%% file: main.tex
\UseRawInputEncoding
\documentclass[journal]{IEEEtran}
\usepackage{pdfpages}

\usepackage{cite}
\usepackage{tikz}
\usetikzlibrary{positioning, calc, backgrounds}
\usepackage{float}
\usepackage{eso-pic}
\usepackage{tabu}
\usepackage{graphicx}
\usepackage{subcaption}
\usepackage[compatibility=false]{caption}

\usepackage{hyperref}

\ifCLASSINFOpdf
\else
\fi
\usepackage{amsmath,amssymb}
\begin{document}

%\input{chapters/front_page}
%\newpage

\title{A Decomposition-Based Framework for Joint Optimization and Spatial Packaging of Interconnected Systems with Physical Interactions}

\author{J.~Bückmann, J. van Kampen, T.~Hofman % <-this % stops a space
%\thanks{Manuscript received Month x, 200x; revised Month x, 200x.}% <-this % stops a space
\thanks{J.~Bückmann, J. van Kampen, and T. Hofman (e-mail: t.hofman@tue.nl) are with the Eindhoven University of Technology (TU/e), Dept. of Mechanical Engineering, \href{https://www.tue.nl/en/research/research-groups/control-systems-technology}{Control Systems Technology} section, \href{https://www.tue.nl/en/research/research-groups/group-hofman}{Engineering Systems Design} group, P.O.Box 513, 5600 MB Eindhoven, The Netherlands.}}

\setcounter{page}{1}

\maketitle

\begin{abstract}
\input{chapters/abstract}

\end{abstract}


% \fi
\IEEEpeerreviewmaketitle

\section{Introduction}
\input{chapters/introduction}

\section{Methods}\label{section:methods}
\input{chapters/methods}

\section{Results}\label{section:Results}
\input{chapters/results}

\section{Discussion}\label{section:discussion}
\input{chapters/discussion}

\section{Conclusion}\label{section:conclusion}

\input{chapters/conclusion}

\section{Future Work}\label{section:Future_work}

\input{chapters/future_work}

% or


% % you can choose not to have a title for an appendix
% % if you want by leaving the argument blank
% \section{}

% % use section* for acknowledgment
\section*{Acknowledgment}
\input{chapters/acknowledgement}

\ifCLASSOPTIONcaptionsoff
  \newpage
\fi



%\bibliographystyle{IEEEtran}
\bibliographystyle{ieeetr}   % or plain, unsrt, IEEEtran, etc.
\bibliography{references}

% \end{IEEEbiography}


%\newpage



%\vfill

\newpage
\appendices
\input{chapters/appendix}

\end{document}

%% file: chapters/abstract.tex
% This paper presents an approach to the optimization of spatial packaging of interconnected systems with physical interactions (SPI2) in three-dimensional component placement problems that improves convergence rate and solution quality through an adapted methodology to enhance numerical robustness for gradient-based optimization and reduce computational load. Existing SPI2 approaches are extended upon through the addition of alignment capabilities to represent port-to-port alignments of components. Furthermore, the applicability of SPI2 is further investigated to allow the physical placement location of components to become an effective design variable, which can be used for penalty coordination to ensure design feasibility and can thus be used in system-level optimization. To that extent, the approach is validated through the use of a multi-objective optimization using Nondominated Sorting Genetic Algorithm II (NSGA-II) with a powertrain optimization sub-problem and battery chassis integration sub-problem to demonstrate the applicability of the SPI2 model in system-level optimization. 
% The results is a two-fold implementation of SPI2 for an automotive use case. First, being used for design generation and second as an integration into a system-level design coordinator capable of outperforming a discretized exhaustive search at a lower computational cost.
This paper presents an approach and application of optimization of spatial packaging of interconnected systems with physical interactions (SPI2) in three-dimensional component placement problems. To enable its application for an automotive use case, SPI2 must support both initial design generation, including component alignment, and robust system-level coordination, requiring improved solution reliability and tractable computational cost.
To address these requirements, the proposed methodology improves convergence rate and solution quality by enhancing numerical robustness in gradient-based optimization while reducing computational load. Existing SPI2 approaches are extended through the addition of alignment capabilities, enabling the representation of port-to-port alignments between components. Furthermore, the applicability of SPI2 is expanded by treating component placement locations as design variables, allowing for penalty-based coordination to ensure design feasibility and enabling integration within system-level optimization.
The approach is validated using a multi-objective optimization framework based on Nondominated Sorting Genetic Algorithm II (NSGA-II), applied to a combined powertrain optimization and battery chassis integration problem. This demonstrates the effectiveness of the SPI2  in a system-level design context.
The results show a twofold application of SPI2 in an automotive use case: first, as a tool for initial design generation, and second, as part of a system-level design coordinator that outperforms a discretized exhaustive search while requiring lower computational cost.

%% file: chapters/introduction.tex
\IEEEPARstart{O}{}ptimization is the cornerstone of technological advancement, driving efficiency, sustainability, and innovation across engineering disciplines. With optimization applications ranging from robot design \cite{Sapietova2018ApplicationDesigning} to packaging problems such as spatial packaging of interconnected systems with physical interactions (SPI2) \cite{Peddada2022TowardSPI2}\cite{Peddada2021AnOpportunities} or vehicle optimization \cite{Othaganont2017Multi-objectiveTopologies}.
Given the rise in computational power, optimization problems can be extended to not only optimize a certain part of a system but also optimize on system-level \cite{Schuman2005IntegratedModeling}. 
In terms of automotive optimization such system and subsystem level optimization strategies are being investigated utilizing problem decomposition strategies \cite{Bayrak2016Decomposition-BasedDesign}. When optimizing at system level, an additional layer of complexity is added, namely the difference in optimization objectives between different subsystems. This means a multi-objective optimization approach must be adopted \cite{Ehrgott2026FiftyComputation}. The extension into multi-objective optimization through problem decomposition strategies, however, leads to extensive design spaces with growing interdependencies to result in a feasible design, and thus coordination methods must be considered to coordinate for feasibility as well as efficient coordination of complex design spaces \cite{Cohen1978OptimizationApproach}. This complexity applies especially to modern technology design that often targets tighter and more compact design layouts\cite{Hofstetter2018Multi-ObjectiveDevelopment}.

\begin{figure}
    \centering
    \includegraphics[width=\linewidth]{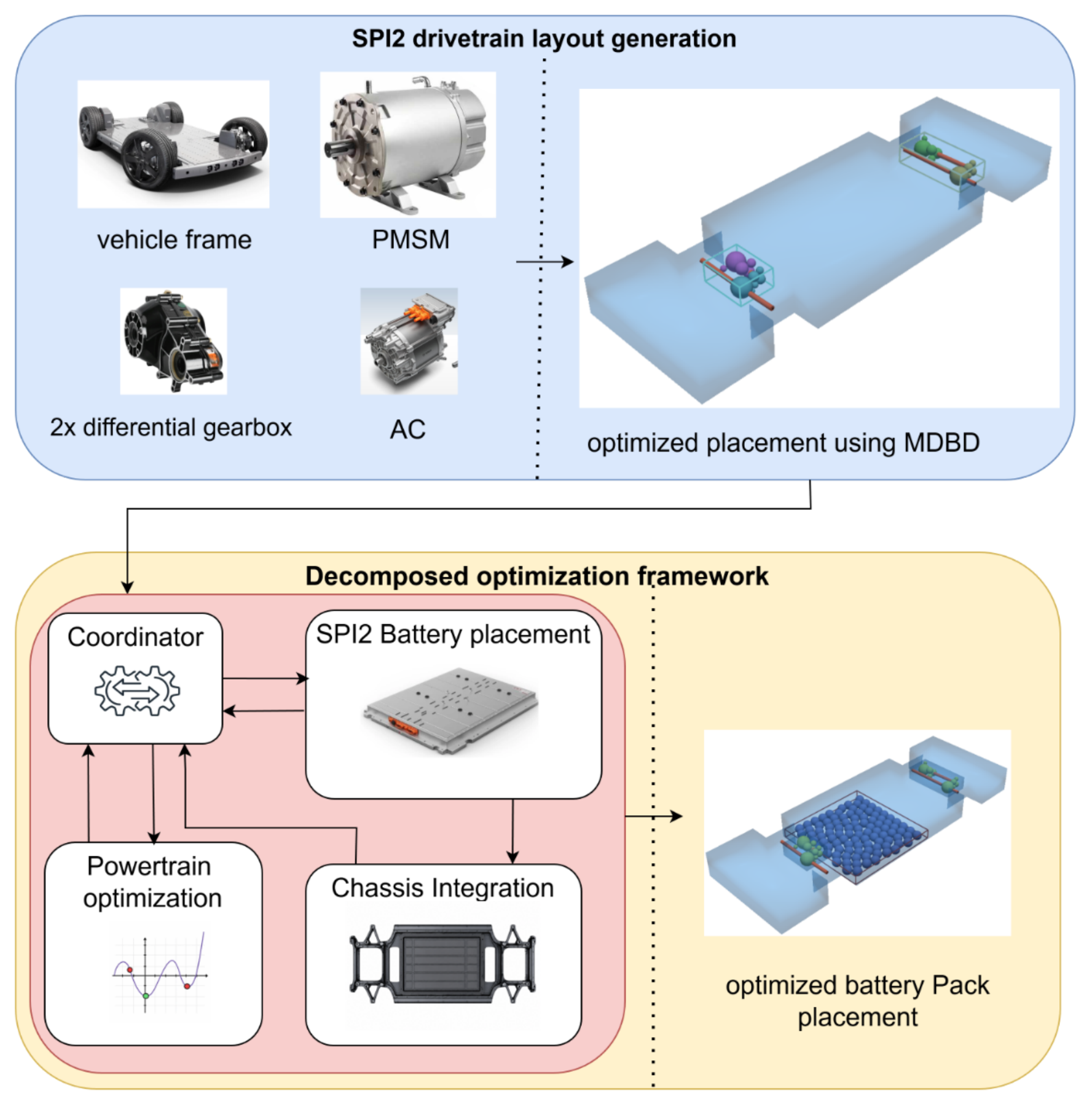}
    \caption{Overview of framework, where MDBD component approximations are used to generate a drivetrain layout and use Analytical Target Cascading as a problem decomposition to optimize placement.}
    \label{fig:workflow_introduction}
\end{figure}

\indent \textit{Related Literature: }
Research efforts address spatial layout and packaging optimization, aiming to minimize volume, ensure feasibility \cite{Hofstetter2018Multi-ObjectiveDevelopment,VanKampen2023}. In an effort to capture component interactions, SPI2 approaches were developed \cite{Peddada2021AnOpportunities}, and the approach was expanded upon to used Maximal Disjoint Ball Decompositions (MDBD)\cite{Chen2022,Behzadi2025SpatialSystems,Westerhof2025HybridSystems.}. 

Parallel research in automotive efficiency optimization primarily focuses on powertrain component sizing and scaling. Electric machine rating, gear ratios, and transmission configurations are commonly treated as design variables in order to shift operating points toward higher-efficiency regions, reduce mass, and satisfy performance constraints \cite{Silvas2016}, or improve handling performance through mass distribution and center of gravity placement \cite{vanKampen2026AutomatedConstraints}.

However, the physical placement of drivetrain components and there scaling through optimization directly influences the longitudinal and vertical position of the vehicle center of gravity. Since the CoG location affects longitudinal load transfer during acceleration and braking, it modifies axle normal forces and, consequently, the admissible traction and regenerative braking forces. Torque allocation strategies in distributed electric drivetrains, which aim to maximize efficiency or stability, explicitly depend on these axle loads \cite{Lin2014AVehicle}. In particular, rear-biased regenerative braking, which is beneficial for energy recuperation, can lead to reduced vehicle stability \cite{Levickas2025StabilityBraking}. As a result, component placement does not merely represent a packaging problem, but can fundamentally affect powertrain utilization, energy recuperation potential, and overall system efficiency. Additionally, modern research presents approaches to battery integration in the vehicle floor as a structural load-bearing element leading to significant mass reduction \cite{Lu2023TopologyBatteries}.
Despite this interdependency, current research that apply decomposition methods for vehicle design largely treats powertrain sizing and component placement as separate optimization problems. The coupling between spatial mass distribution and energy-optimal powertrain operation is rarely modelled within a unified framework. This separation neglects potentially beneficial trade-offs between packaging decisions and drivetrain efficiency.
A likely cause being the complexity of the problem. Therefore, problem decompositions are implemented. These approaches have been successfully applied in complex vehicle design problems, including drivetrain and transmission optimization using Analytical Target Cascading (ATC) \cite{Fahdzyana2022,Kim2003AnalyticalDesign}. However, research integrating placement into these decomposed problems is uncommon.

To the authors’ knowledge, the explicit integration of simultaneous component scaling, battery chassis integration effects and physically coupled spatial placement within a unified hierarchical optimization framework has not previously been reported for vehicle system design.

\indent \textit{Statement of Contribution: }
To address this gap, the present work proposes a joint optimization framework that integrates component placement, component integration into the vehicle structure, and powertrain scaling within a unified system-level formulation.  Therefore, a decomposition strategy is employed to partition the overall problem into interacting subproblems while preserving coordination toward a system-level optimum.
% This work presents a hierarchical optimization framework that integrates spatial packaging considerations into a decomposed powertrain optimization problem in which interacting subsystems are scaled and coordinated simultaneously. 
Rather than treating physical placement as a downstream feasibility check, the proposed formulation incorporates spatial positioning as an explicit design variable that interacts with component scaling and system performance. The intended Framework is presented in Fig. \ref{fig:workflow_introduction}.

A single powertrain architecture is considered, and its components are optimized through continuous scaling variables and transmission parameters. In contrast to conventional approaches where component sizing and packaging are addressed sequentially, the present framework embeds spatial placement directly within the optimization process. The placement of a primary subsystem influences system-level behavior through physically coupled mechanisms, thereby introducing non-trivial interactions between mass distribution, structural compostion, and performance metrics.

% The coordination is achieved using an ATC inspired formulation in which spatial variables are treated as coupling quantities between subsystems. A system-level coordinator iteratively adjusts placement targets, while subsystem-level optimizations determine the optimal component scaling and integration of components into the vehicle structure. This structure preserves subsystem autonomy, enables robust handling of nonlinear and potentially non-smooth behavior, and allows interacting physical systems to be optimized in a consistent hierarchical manner.

The primary contribution of this work is therefore methodological: it demonstrates how spatial packaging of interacting mechanical systems can be incorporated to make positioning an active optimization variable within a decomposed architecture, rather than being treated as a geometric constraint imposed after performance optimization. This integration enables systematic exploration of the coupled effects of component scaling and physical placement on overall system performance and establishes a methodology for jointly addressing parametric and spatial design decisions.

\indent \textit{Paper Layout: } The paper will present the performed research with Section \ref{section:methods} covering the applied methodology, Section \ref{section:Results} covers the results, Section \ref{section:discussion} discusses the findings Section, \ref{section:conclusion} presents a conclusion and Section \ref{section:Future_work} provides an outlook on future research.

%% file: chapters/methods.tex
% this section contains mehtods used so make sure there is a reference to Stevens' paper for existing contraint definitions

% make sure there is a "catchy plaatje" that shows all steps in taken in the full opimization
This section presents the methodology used to formulate and solve the integrated system-level optimization problem. First, the SPI2 is introduced, including several methodological adaptations to existing approaches.
Subsequently, the subproblems for a demonstrative decomposed system level optimizing problem are presented. The first subproblem consists of a standalone powertrain optimization routine, in which component scaling and transmission parameters are optimized while maintaining dependencies on system-level variables such as mass distribution and component placement. The second subproblem describes an analytical model for battery integration into the vehicle chassis, capturing the effects of spatial placement on structural stiffness and mass through a reduced-order representation.
Finally, the multi-objective optimization framework is presented. This includes the coordination strategy used to couple the subproblems, and the role of SPI2 as a mechanism to enforce spatial feasibility and enable consistent system-level design exploration within the optimization process.

\subsection{Placement Problem SPI2}\label{section:placement}
\subsubsection{Model Setup prerequisites}
The placement problem is formulated as an extension of the methodology proposed by Westerhof \cite{Westerhof2025HybridSystems.} and Behzadi \cite{Behzadi2025SpatialSystems} employing the described constraint equations for routing–routing, routing–object, and object–object interference. Utilizing the MDBD defined by Chen \cite{Chen2020MaximalAnalysis}.

In the present work, several methodological extensions are introduced to enhance numerical robustness, geometric generality, and solver compatibility with large-scale nonlinear programming. 

First, the original Euler-angle–based rotation parameterization is replaced by a quaternion-based rotation representation. This modification eliminates singularities (gimbal lock), removes axis-order dependencies, and ensures smooth differentiability of orientation variables within the nonlinear programming framework.

Second, port-alignment constraints for routing are incorporated to ensure physically consistent directional coupling between component interfaces and connected routing segments. These constraints enforce alignment between predefined port normal vectors and routing direction vectors, thereby guaranteeing feasible connection geometries during optimization.

Third, a signed distance field (SDF)–based boundary constraint formulation is implemented to enforce enclosure feasibility. Instead of penalty-based boundary handling, a precomputed volumetric signed distance representation of the admissible domain is embedded symbolically into the optimization problem, enabling smooth and differentiable inequality constraints for all sphere-based object representations.

\subsubsection{Geometric Representation of Objects and Ports}

To formally describe the placement problem, each component is represented as a rigid object composed of primitive geometric elements and connection interfaces.

Let $A_i$ denote object $i$, where $i \in \{1, \dots, n_{\text{obj}}\}$ uniquely identifies each object in the system.

Each object $A_i$ is represented by a set of spheres obtained from a maximal disjoint ball decomposition (MDBD). The set of spheres associated with object $i$ is defined as
\begin{equation}
\mathcal{S}_i = \left\{ (c_{i,j}, r_{i,j}) \mid j = 1, \dots, n_i^{\text{sph}} \right\},
\end{equation}
where $j$ indexes the spheres belonging to object $i$. Each sphere is defined by a center point $c_{i,j} \in \mathbb{R}^3$ and a radius $r_{i,j} > 0$. All sphere centers $c_{i,j}$ are expressed in the local coordinate frame $\mathbb{F}_{A_i}$ of object $A_i$.

In addition to its geometric representation, each object may contain a set of connection ports used to define routing interfaces. The set of ports on object $A_i$ is defined as
\begin{equation}
\mathcal{P}_i = \left\{ p_{i,\ell} \mid \ell = 1, \dots, n_i^{\text{port}} \right\},
\end{equation}
where $\ell$ indexes the ports on object $i$, and each port location $p_{i,\ell} \in \mathbb{R}^3$ is expressed in the same local frame $\mathbb{F}_{A_i}$.

Optionally, each port may be associated with a predefined local direction vector $d^{\text{loc}}_{i,\ell} \in \mathbb{R}^3$, specifying the required orientation of a connection at that interface.

\begin{figure}[t]
    \centering
    \includegraphics[width=\linewidth]{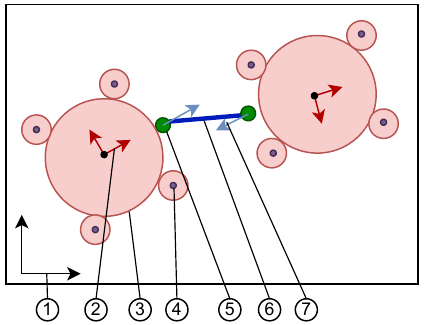}
    \caption{Depiction of components in model: Workspace frame $\mathbb{F_W}$ (1, White), Object frame $\mathbb{F}_{A_i}$ (2, Dark Red), Object spheres $\mathcal{S}_i$ (3, Light Red), Sphere centre point $c_{i,j}$ (4, Purple), Port $\mathcal{P}_i$ (5, Green), Routing $L$ (6, Dark Blue), Required port direction vector $d^{\text{loc}}_{i,\ell}$ (7, Light Blue)}
    \label{fig:model_depiction}
\end{figure}

Given a rigid-body transformation consisting of a rotation matrix $R \in SO(3)$ and a translation vector $t_i \in \mathbb{R}^3$, all geometric entities of object $A_i$ are mapped to the workspace frame $\mathbb{F_W}$ as
\begin{equation}
p^W = R(q_i) p + t_i,
\end{equation}
where $p$ denotes any point defined in the local frame $F_{A_i}$, including sphere centers $c_{i,j}$ and port locations $p_{i,\ell}$. This transformation fully defines the spatial placement of object $A_i$ in the workspace.
These component definitions are visually presented in Fig \ref{fig:model_depiction}.

These extensions preserve the core MDBD-based interference formulation of Westerhof \cite{Westerhof2025HybridSystems.} and Behzadi \cite{Behzadi2025SpatialSystems} while improving numerical stability, geometric expressiveness, and compatibility with gradient-based solvers such as IPOPT.

In the present work, several modeling choices are adapted to reflect the mechanical drivetrain interpretation of routing.

First, routing segments represent mechanical axles which must adhere to alignment constraints, connections such as electric cabling and cooling routing are assumed to have sufficient flexibility in design to be routed around components in later design stages as a chain of connected routing segments with intermediate nodes and are therefore omitted. Consequently, a routing $L$ is restricted to a single straight segment. Instead of allowing a chain of connected routing segments as in \cite{Westerhof2025HybridSystems.}, it directly connects two ports $p_{i,k}$ and $p_{j,\ell}$. No intermediate nodes or routing control points are introduced. The routing geometry is therefore completely defined by the transformed positions of its two associated ports in the workspace frame.

Second, the rigid-body pose of each object $i$ is parameterized using a quaternion-based representation, which is formally presented here to fully define the action space. The reasoning for implementation and methodology will follow later. The orientation of object $A_i$ is defined by four decision variables
\begin{equation}
    q_i = [w_i,\, q_{1,i},\, q_{2,i},\, q_{3,i}]^\top.
\end{equation}
where the scalar $w_i$ and vector part $q_{\cdot ,i}$ are the rotational decision variables.
% which are normalized prior to evaluation to ensure $\|q_i\| = 1$.
The translation $t$ of object $i$ in the workspace frame $x$, $y$ and $z$ axis is defined by
\begin{equation}
    t_i = [x_i,\, y_i,\, z_i]^\top \in \mathbb{R}^3.
\end{equation}
Hence, each movable object contributes seven decision variables, and its pose is expressed as
\begin{equation}
    x_{A_i} =
    \begin{bmatrix}
        q_i^\top & t_i^\top
    \end{bmatrix}^\top
    \in \mathbb{R}^7.
\end{equation}

% The quaternion is mapped to a rotation matrix $R(q_i) \in SO(3)$, and the rigid transformation from object frame $\mathbb{F}_{A_i}$ to workspace frame $\mathbb{F_W}$ becomes
% \begin{equation}
%     p^W = R(q_i)\, p^{A_i} + t_i.
% \end{equation}

Third, the first object, representing the connection ports location of the wheels in $\mathbb{F_W}$, is fixed and therefore does not contribute decision variables. This object contains the port locations corresponding to the wheel--boundary interface of the drivetrain layout. Since these ports define the mechanical interface to the enclosing boundary and must remain fixed in space, object $1$ is locked in the workspace frame:
\begin{equation}
    R_1 = I, \qquad t_1 = 0.
\end{equation}

Consequently, the total number of design variables becomes
\begin{equation}
    n_{\text{var}} = 7 \left(n_{\text{obj}} - 1\right),
\end{equation}
as only objects $i = 2, \dots, n_{\text{obj}}$ are free to move.

The complete design vector is therefore written as
\begin{equation}
    \mathbf{x} =
    \begin{bmatrix}
        x_{A_2}^\top & \dots & x_{A_{n_{\text{obj}}}}^\top
    \end{bmatrix}^\top.
\end{equation}

\subsubsection{Quaternion-Based Rotation Representation}
An accurate representation of rigid-body orientation is essential for the placement optimization. In this work, rotation is parameterized using unit quaternions \cite{Shoemake1985AnimatingCurves}, which provide a globally non-singular representation of 3-D rotational motion. 
This replaces the Roll-Pitch-Yaw parameterization used in previous works\cite{Westerhof2025HybridSystems.}\cite{Behzadi2025SpatialSystems}, in an effort to eliminate singularities and improve the smoothness of the nonlinear program.

Euler-angle parameterizations are known to be locally minimal but suffer from intrinsic singularities due to the topology of the rotation group $SO(3)$. In particular, gimbal lock occurs when two rotation axes align, resulting in a loss of one degree of freedom. Furthermore, Euler representations exhibit axis-order dependence, discontinuities and singularities \cite{Pavllo2019ModelingNetworks}. These effects introduce non-smoothness into objective and constraint functions, which can impair convergence of gradient-based optimization algorithms.

A quaternion
\begin{equation}
q = [w,\, q_1,\, q_2,\, q_3]^{\mathsf{T}},
\end{equation}

represents a rotation of angle $\theta$ about a unit axis $\mathbf{u}$.
% through
% \[
% w = \cos\left(\frac{\theta}{2}\right), \qquad (x,y,z) = \mathbf{u}\,\sin\left(\frac{\theta}{2}\right).
% \]
The set of unit quaternions forms the 3-sphere $S^{3}$, which provides a smooth and globally non-singular double cover of $SO(3)$. The quaternion formulation and its computational advantages for rotation interpolation and numerical robustness were established by Shoemake \cite{Shoemake1985AnimatingCurves}. The absence of coordinate singularities ensures that composed geometric mappings remain continuously differentiable with respect to the decision variables.

To utilize the unit quaternion for rotation, a normalization strategy is applied. Let $q_{\mathrm{d}} \in \mathbb{R}^4$ denote the unconstrained quaternion parameter vector used as optimization decision variable. To avoid introducing an explicit nonlinear equality constraint $\|q\|=1$, the quaternion components are optimized in $\mathbb{R}^4$ and subsequently normalized before evaluation. The normalized quaternion is obtained as
\begin{equation}
    \tilde{q} = \frac{q_{\mathrm{d}}}{\|q_{\mathrm{d}}\| + \varepsilon}.
\end{equation}
This strategy follows the normalization approach described by Sola \cite{Sola2017QuaternionFilter}, where quaternion updates are performed in $\mathbb{R}^4$ and subsequently projected onto the unit sphere $S^3$. The normalization map is smooth for $q \neq 0$, and the small regularization term $\varepsilon$ prevents numerical singularities while preserving differentiability. 

The rotation matrix $R$ used in all geometric computations is obtained from the normalized quaternion $\tilde{q} = (w,x,y,z)$ using the standard closed-form expression defined by Diebel \cite{Diebel2006RepresentingVectors}
% \begin{equation}
%     R(\tilde{q}) =
% \begin{bmatrix}
% 1 - 2(y^{2} + z^{2}) & 2(xy - wz)           & 2(xz + wy) \\
% 2(xy + wz)           & 1 - 2(x^{2} + z^{2}) & 2(yz - wx) \\
% 2(xz - wy)           & 2(yz + wx)           & 1 - 2(x^{2} + y^{2})
% \end{bmatrix}.
% \end{equation}
\begin{equation}
    R(\tilde{q}) =
\begin{bmatrix}
1 - 2(q_2^{2} + q_3^{2}) & 2(q_1q_2 - wq_3)           & 2(q_1q_3 + wq_2) \\
2(q_1q_2 + wq_3)           & 1 - 2(q_1^{2} + q_3^{2}) & 2(q_2q_3 - wq_1) \\
2(q_1q_3 - wq_2)           & 2(q_2q_3 + wq_1)           & 1 - 2(q_1^{2} + q_2^{2})
\end{bmatrix}.
\end{equation}

\subsubsection{Signed Distance Field-Based Boundary Constraint}
In order to obtain a feasible design, components must be placed in the feasible design space.
To enforce non-penetration of all placed components with respect to any
arbitrary enclosing geometry, a signed distance field (SDF) representation of the admissible boundary volume is employed. 
% The distance is first discretized on a structured grid and evaluated using interpolation \cite{Pujol2023AdaptiveInterpolation}. The use of interpolation allows for a smooth approximation of the distance to the surface through quick interpolation.
% opposed to an expensive nearest neighbour search to every mesh query point.

% \paragraph*{SDF construction}

% Let $\mathcal{M}$ denote a closed triangular surface mesh describing
% the admissible boundary volume. For a query point $\mathbf{p} \in \mathbb{R}^3$
% and a surface point $\mathbf{q}$ on the mesh boundary
% $\partial \mathcal{M}$, A signed distance field
% $\phi : \mathbb{R}^3 \rightarrow \mathbb{R}$ is defined as

% \begin{equation}
% \phi(\mathbf{p}) =
% \begin{cases}
% \phantom{-}\min_{\mathbf{q} \in \partial \mathcal{M}}
% \|\mathbf{p}-\mathbf{q}\|,
% & \mathbf{p} \in \mathcal{M}, \\[6pt]
% -\min_{\mathbf{q} \in \partial \mathcal{M}}
% \|\mathbf{p}-\mathbf{q}\|,
% & \mathbf{p} \notin \mathcal{M},
% \end{cases}
% \end{equation}

% The SDF is sampled on a structured Cartesian grid with resolution
% $(n_x,n_y,n_z)$ over a rectangular workspace domain of dimensions
% $(L_x,L_y,L_z)$. The corresponding grid spacings are

% \begin{equation}
% h_x = \frac{L_x}{n_x - 1}, \quad
% h_y = \frac{L_y}{n_y - 1}, \quad
% h_z = \frac{L_z}{n_z - 1}.
% \label{gridspacing}
% \end{equation}

% Signed distances are computed using mesh proximity queries
% combined with inside–outside classification. 
\paragraph*{Signed Distance Field Construction}
Let $\mathcal{M}$ denote a closed triangular surface mesh
describing the admissible boundary volume. For a query point
$\mathbf{p} \in \mathbb{R}^3$ and a surface point
$\mathbf{q}$ on the mesh boundary $\partial \mathcal{M}$,
the signed distance field (SDF)
$\phi : \mathbb{R}^3 \rightarrow \mathbb{R}$ is defined as

\begin{equation}
\phi(\mathbf{p}) =
\begin{cases}
\phantom{-}\displaystyle
\min_{\mathbf{q}\in\partial\mathcal{M}}
\|\mathbf{p}-\mathbf{q}\|,
& \mathbf{p}\in\mathcal{M}, \\[8pt]
-\displaystyle
\min_{\mathbf{q}\in\partial\mathcal{M}}
\|\mathbf{p}-\mathbf{q}\|,
& \mathbf{p}\notin\mathcal{M}.
\end{cases}
\end{equation}

\begin{figure}
    \centering
    \includegraphics[width=\linewidth]{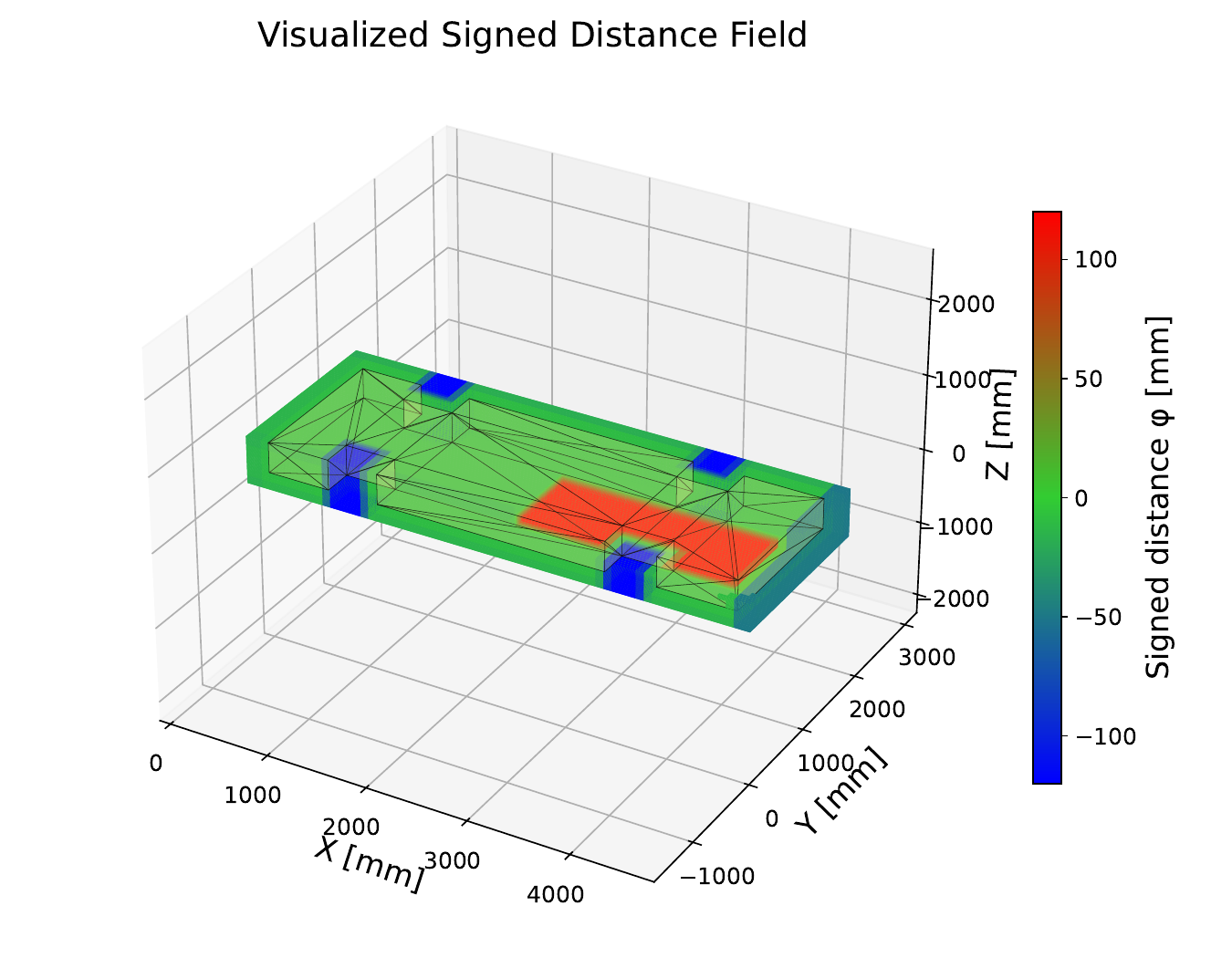}
    \caption{Smooth SDF constructed using cubic B-splines of the vehicle use case on which 3D query-points can be sampled. The top corner has been removed to showcase part of the internal field.}
    \label{fig:SDF}
\end{figure}
\paragraph*{Cubic B-spline Interpolation}

The SDF itself is constructed using cubic B-splines of degree $n=3$ the interpolation is $C^{n-1}=C^2$ continuous
within the sampled domain. Consequently, the field provides
continuous function values and first derivatives, ensuring
smooth boundary evaluation compatible with gradient-based
nonlinear optimization \cite{Yang2022AAccuracy}.

The sampled signed distance field is converted into a smooth volumetric representation using a cubic B-spline interpolant

\begin{equation}
\phi_{\text{spline}}(p)=
\sum_i \sum_j \sum_k
\phi_{ijk}\, B_i(x)\,B_j(y)\,B_k(z),
\end{equation}

where $i$, $j$, and $k$ index the grid nodes along the $x$-, $y$-, and
$z$-directions, and
$\phi_{ijk} := \phi(x_i,y_j,z_k)$ denote the signed distance values
sampled on the Cartesian grid. The cubic B-spline basis functions
$B(\cdot)$ follow the standard formulation of de Boor\cite{R.1980ASplines.}. The resulting smooth signed distance field on which points can be sampled is shown in Fig. \ref{fig:SDF}.

\paragraph*{Boundary Constraint Formulation}
All components are represented as unions of spheres.
Let $\mathbf{c}_{i,j} \in \mathbb{R}^3$ denote the center of the
$j$-th sphere of object $i$, and let $r_{i,j} > 0$ denote its radius.
For each sphere, the signed clearance margin is defined as

\begin{equation}
g_{i,j} =
\phi_{\text{spline}}(\mathbf{c}_{i,j})
- r_{i,j}.
\end{equation}

Boundary feasibility is enforced through the hard inequality constraint

\begin{equation}
g_{i,j} \ge 0
\quad \forall i,j.
\end{equation}

% \paragraph*{Constraint Evaluation}

% All boundary constraints are appended to the global constraint
% vector of the nonlinear program. During optimization,
% $\phi_{\text{world}}$ and its gradients are evaluated symbolically
% at each iteration.

% The use of tricubic interpolation provides smooth implicit
% boundary representation with continuous gradients and higher-order
% approximation accuracy, enabling reliable geometric feasibility
% enforcement in large-scale non-convex placement problems.

% \subsection{Objective}

% \subsection{Constraints}
% %SDF om smooth contraint te behouden,

\subsubsection{Port-Alignment Constraints for Routing}
\label{sec:alignment_constraints}

In addition to collision-free routing, the synthesized connection paths must satisfy port-direction requirements: each routed connection should leave and enter a component through a prescribed outward direction at the corresponding port. This is required to respect physical connector orientations, defining in which direction a mechanical axle exits or enters a component, and avoid routing solutions that are geometrically feasible but practically unrealizable.

To formally define the actual routing direction at each
connected port, the direction vector is derived directly
from the two connected port locations in the workspace
frame. Since each routing $L$ is restricted to a single
straight segment ($K_L = 1$), its geometry is completely
defined by the transformed positions of its two associated
ports.

Let port $(i,\pi)$ of object $i$ and port $(j,\rho)$ of
object $j$ be connected by routing $k$. Let
$\mathbf{p}^{W}_{i,\pi}$ and $\mathbf{p}^{W}_{j,\rho}$
denote their positions in the workspace frame. These are
obtained from the rigid-body transformations

\begin{equation}
\mathbf{p}^{W}_{i,\pi}
=
R_i \mathbf{p}^{A_i}_{i,\pi} + \mathbf{t}_i,
\end{equation}

\begin{equation}
\mathbf{p}^{W}_{j,\rho}
=
R_j \mathbf{p}^{A_j}_{j,\rho} + \mathbf{t}_j,
\end{equation}

where $R_i$ and $R_j$ are the rotation matrices derived
from the quaternion parameterization and
$\mathbf{t}_i, \mathbf{t}_j$ are the object translations.

The routing direction at the start port $(i,\pi)$ is defined
as the normalized vector pointing from that port toward
the connected port:

\begin{equation}
\mathbf{d}^{(k)}_{i,\pi}
=
\frac{
\mathbf{p}^{W}_{j,\rho}
-
\mathbf{p}^{W}_{i,\pi}
}{
\left\|
\mathbf{p}^{W}_{j,\rho}
-
\mathbf{p}^{W}_{i,\pi}
\right\|
+
\varepsilon
}.
\label{eq:dir_start}
\end{equation}

Analogously, the direction at the receiving port $(j,\rho)$
is defined as

\begin{equation}
\mathbf{d}^{(k)}_{j,\rho}
=
\frac{
\mathbf{p}^{W}_{i,\pi}
-
\mathbf{p}^{W}_{j,\rho}
}{
\left\|
\mathbf{p}^{W}_{i,\pi}
-
\mathbf{p}^{W}_{j,\rho}
\right\|
+
\varepsilon
}.
\label{eq:dir_end}
\end{equation}

Here $\varepsilon$ is a small regularization constant that
ensures differentiability and prevents division by zero in
degenerate configurations.

For each port $(i,\pi)$ (object index $i$, port index $\pi$), a desired direction is specified in the local
object frame as $\mathbf{d}^{\mathrm{loc}}_{i,\pi} \in \mathbb{R}^3$ typically a unit vector.
Given the object rotation matrix $R_i$ (obtained from the quaternion parameterization used in the placement model),
the desired direction in the world frame is
\begin{equation}
\mathbf{d}^{\mathrm{w}}_{i,\pi} =
\frac{R_i \, \mathbf{d}^{\mathrm{loc}}_{i,\pi}}
{\lVert R_i \, \mathbf{d}^{\mathrm{loc}}_{i,\pi}\rVert+\varepsilon}.
\label{eq:dir_des_world}
\end{equation}
Normalizing both desired and actual directions ensures that their dot product directly equals the cosine of the
misalignment angle.

Let $\theta$ denote the angle between the actual direction $\mathbf{d}_{\mathrm{act}}$ and desired world direction
$\mathbf{d}^{\mathrm{w}}$ at a port. The cosine similarity is
\begin{equation}
c = \left(\mathbf{d}^{\mathrm{w}}\right)^\top \mathbf{d}_{\mathrm{act}} = \cos(\theta).
\label{eq:cosine_similarity}
\end{equation}
A maximum admissible misalignment margin $\theta_{\max}$ is prescribed. This can be expressed as
a smooth inequality using a cosine threshold $c_{\mathrm{thr}} = \cos(\theta_{\max})$:
\begin{equation}
g_{\mathrm{dir}} = c - c_{\mathrm{thr}} \ge 0.
\label{eq:hard_alignment_constraint}
\end{equation}
Constraint~\eqref{eq:hard_alignment_constraint} is enforced for both ends of every routed connection for which a desired
direction is defined. Geometrically, it restricts the routed path to remain within a cone about the desired outward
direction $\theta_{\max}$.

While the hard constraint ensures feasibility, it may lead to poor conditioning when the optimizer explores infeasible
regions. Therefore, a smooth penalty is added to the objective to guide the solver toward
strong alignment and to provide informative gradients outside the feasible cone.

An always-active term encourages $\theta \rightarrow 0$ even when the solution already satisfies the margin:
\begin{equation}
J_{\mathrm{align}} = n_{align} - c.
\label{eq:align_pull}
\end{equation}
where $n_{align}$ is the number of alignment constraints such that the minimum value of the alignment objective is limited to 0.
% Using $1-c$ is convenient because for small angles $1-\cos(\theta) \approx \theta^2/2$, providing a locally quadratic, well-scaled penalty.

% \paragraph{Violation penalty with smooth hinge.}
% To penalize only margin violations smoothly, a softplus-based hinge is used:
% \begin{equation}
% J_{\mathrm{viol}} =
% \mathrm{softplus}\!\left(c_{\mathrm{thr}} - c\right),
% \qquad
% \mathrm{softplus}(z)=\frac{1}{\alpha}\ln\!\left(1+e^{\alpha z}\right),
% \label{eq:softplus_violation}
% \end{equation}
% where $\alpha>0$ controls the sharpness. For $c \ge c_{\mathrm{thr}}$, the argument is non-positive and the penalty
% approaches zero smoothly; for $c < c_{\mathrm{thr}}$, it grows approximately linearly with $(c_{\mathrm{thr}}-c)$.

% \paragraph{Total soft term.}
% For each constrained port direction, the combined contribution is
% \begin{equation}
% J_{\mathrm{dir}} = w_{\mathrm{align}}(1-c) + w_{\mathrm{viol}}\,\mathrm{softplus}(c_{\mathrm{thr}}-c),
% \label{eq:total_soft_alignment}
% \end{equation}
% and the global alignment penalty is the sum over all constrained ports of all routed connections.

\subsubsection{objective}
The objective function of the placement problem is defined as a weighted summation of the separate objectives to obtain a minimized objective value $J$,

\begin{equation}
    J = w_{vol}J_{vol}+w_{align}J_{align}+w_{CoG}J_{CoG},
\end{equation}

where $w$ indicate the weights and $J$ indicated the objective value, $J_{vol}$ and $J_{CoG}$ are implemented as described by Westerhof \cite{Westerhof2025HybridSystems.}.

% \subsection{Powertrain optimization}
% % introduce the powertrain optmization problem and 

% \subsubsection{optimization strategy}
% % define how and why the optimization strategy is approached the way it is such that it can later be reason that this is a possible way to integrate it into the ATC later

% \subsubsection{optimization problem}

% \paragraph*{objective}
% % define cost function

% \paragraph*{constraints}
% % define here how every contraint is checked

\subsection{Powertrain Optimization}\label{section:powertrain}
This section of the methods describes a standalone powertrain optimization routine with design dependencies on the mass and mass distribution of SPI2 results. that will result in an objective value and a response in mass redistribution from component scaling to the system level coordinator.

\subsubsection{Drivetrain Topology}

The investigated vehicle configuration corresponds to a commonly used dual-motor all-wheel-drive (AWD) electric vehicle architecture consisting of a rear-axle permanent magnet synchronous machine (PMSM) and a front-axle alternating current (AC) machine.  
Both machines are connected to their respective axles via fixed single-stage gear reductions. 
Such front–rear dual-motor architectures are widely adopted in modern battery electric vehicles due to their modularity, redundancy, and controllability.

The rear axle is driven by the PMSM through gear ratio $\gamma_{\mathrm{rear}}$, while the front axle is driven by the AC machine through gear ratio $\gamma_{\mathrm{front}}$. 
Both machines are continuously scalable through dimensionless scaling factors
\begin{equation}
s_{\mathrm{pmsm}} > 0,
\qquad
s_{\mathrm{ac}} > 0,
\end{equation}
which proportionally scale the rated torque capability of the base machine models.

% In the present formulation, the topology is fixed and no discrete architectural decisions are optimized. 
% The design variables are therefore limited to machine scaling and transmission parameters.

\subsubsection{Problem Statement}

For the center-of-gravity position $x_{\mathrm{cg}}$ and $z_{\mathrm{cg}}$ and system mass $m$, provided by the system-level coordinator, the powertrain subsystem solves the parametric optimization problem
\begin{equation}
\min_{s_{\mathrm{pmsm}},\,s_{\mathrm{ac}},\,\gamma_{\mathrm{rear}},\,\gamma_{\mathrm{front}}}
J_{\mathrm{pt}}(s_{\mathrm{pmsm}},s_{\mathrm{ac}},\gamma_{\mathrm{rear}},\gamma_{\mathrm{front}}; x_{cg},z_{cg},m)
\end{equation}
subject to vehicle performance constraints.

The objective corresponds to the net electrical energy consumption over a driving cycle.

\subsubsection{WLTP-Based Longitudinal Demand Model}

The longitudinal vehicle dynamics \cite{Guzzella2007VehicleOptimization} are evaluated over the WLTP Class~3 cycle \cite{Tutuianu2015DevelopmentLegislation}. 
Let $v_k$ denote the discrete velocity samples and $\Delta t_k$ the corresponding time increments. 
The longitudinal acceleration is computed as
\begin{equation}
a_k = \frac{v_k - v_{k-1}}{\Delta t_k}.
\end{equation}

The total tractive force demand at the wheels is
\begin{equation}
F_x(k) =
\frac{1}{2} \cdot\rho\cdot C_d \cdot A\cdot v_k^2
+ m \cdot g \cdot C_{rr}
+ m \cdot a_k,
\end{equation}
where $m$ denotes the total vehicle mass, $\rho$ the air density, $C_d$ the drag coefficient, $A$ the frontal area, $g$ the gravitaional constant, and $C_{rr}$ the rolling resistance coefficient. 
The corresponding wheel torque $T_w$ is
\begin{equation}
T_w(k) = F_x(k) \cdot r_w,
\end{equation}
with wheel radius $r_w$.

\subsubsection{Axle Force Allocation}
To avoid deriving an optimal powersplit optimization for the sake of this research, but still have a dependency on mass distribution in the net energy, a physics-based powersplit is used. The ratio of longitudinal braking force on each axle is determined by the dynamic brake force distribution. The distribution is defined by the ratio $\beta$ of vertical load on the front axle $F_{z,f}$ and rear axle $F_{z,r}$ as 
\begin{equation}
    \beta = \frac{F_{z,f}}{F_{z,r}+F_{z,f}}.
\end{equation}
The ratio varies based on the mass distribution of the vehicle and the vehicle's longitudinal acceleration, and is obtained through
\begin{equation}
F_{z,f} = m\cdot g \cdot\left(1 - \frac{x_{\mathrm{cg}}}{L}\right) - \frac{m\cdot a_x\cdot z_{cg}}{L},
\end{equation}
\begin{equation}
F_{z,r} = m\cdot g\cdot \frac{x_{\mathrm{cg}}}{L} + \frac{m\cdot a_x\cdot z_{cg}}{L}.
\end{equation}
where $L$ denotes the wheelbase. 

During normal driving, the powersplit between front and rear axle is considered 1:1. This implementation leaves us with a dependency on mass distribution for energy use without requiring the additional integration of an additional powersplit control level optimization. Meaning that this optimization problem is set up with sufficient dependencies on component placement and mass distribution without too much complexity to use as a demonstration for SPI2 integration into system-level design.
\subsubsection{Electric Machine and 
Transmission Model} 
Each motor denoted by  $i \in {front,rear}$ speeds are related to vehicle velocity through the fixed gear ratios:
\begin{equation}
\omega_{i}(k) = \frac{v_k \cdot\gamma_{i}}{r_w},
\end{equation}
Motor torques follow from axle force allocation for $F_i (k)\ge0$
\begin{equation}
T_{i}(k) = \frac{1}{2}\cdot\frac{F_{x}(k)\cdot r_w}{\gamma_{i}},
\end{equation}
meaning half the driving force is from the front axle and half the driving force is from the rear axle
and for $F_i (k)<0$
\begin{equation}
T_{front}(k) = \beta\cdot \frac{F_{x}(k)\cdot r_w}{\gamma_{i}},
\end{equation}
\begin{equation}
T_{rear}(k) = (1-\beta)\cdot \frac{F_{x}(k)\cdot r_w}{\gamma_{i}},
\end{equation}
The rated torque envelopes of the motors scale linearly with the scaling factors:
\begin{equation}
T_{\max}^{\mathrm{rear}} = s_{\mathrm{pmsm}} \cdot T_{\max}^{\mathrm{rear,base}},
\qquad
T_{\max}^{\mathrm{front}} = s_{\mathrm{ac}} \cdot T_{\max}^{\mathrm{front,base}},
\end{equation}
and the corresponding efficiency maps $\eta_{i,map}$ maps the efficiency to the new torque scaling.

Machine efficiency is evaluated through two-dimensional interpolation of a set of efficiency maps,
\begin{equation}
\eta_i = \eta_{i,map}(\omega_i, T_i).
\end{equation}
Mechanical output power $P_{i,out}$ is 
\begin{equation}
    P_{i,out} = \omega_i\cdot T_i
\end{equation}
Electrical input power $P_{i,in}$ is computed as
\begin{equation}
P_{i,\mathrm{in}} =
\begin{cases}
\displaystyle \frac{P_{i,\mathrm{out}}}{\eta_i}, & T_i \ge 0, \\
P_{i,\mathrm{out}} \cdot\eta_i, & T_i < 0.
\end{cases}
\end{equation}
% \begin{equation}
% P_{\mathrm{in}}(k) =
% \begin{cases}
% \displaystyle \frac{P_{\mathrm{out}}(k)}{\eta(k)}, & T(k) \ge 0, \\
% P_{\mathrm{out}}(k) \cdot\eta(k), & T (k)< 0.
% \end{cases}
% \end{equation}
The cycle energy consumption is
\begin{equation}
E_{\mathrm{in}} =
\sum_i\sum_k P_{i,\mathrm{in}}(k)\cdot\Delta t_k.
\end{equation}
The subsystem objective is therefore defined as
\begin{equation}
J_{\mathrm{pt}} = E_{\mathrm{in}}.
\end{equation}

% where the Energy regeneration $E_{regen}$  is determined through the brake force distribution and per axle, and the corresponding efficiency maps for the motor on each axle

% The resulting normal forces define tire force limits
% \begin{equation}
% |F_{x,i}| \le \mu F_{z,i}.
% \end{equation}

% This dynamic brake force distribution model ensures physically admissible force allocation while maintaining stable vehicle handling.

\subsubsection{Constraints}

The following performance constraints are enforced to obtain a well-defined design space for the component scaling problem \cite{Verbruggen2020ElectricTrucks}.

\paragraph{Top Speed Constraint $g_{pt,1}$}
\begin{equation}
\min
\left(
\frac{r_w \cdot\omega_{i,\max}}{\gamma_{i}}
\right)
\ge v_{\mathrm{req}}.
\end{equation}
where $v_{req}$ is the required tops speed and $\omega_{i,max}$ is the maximum rated motor speed. To validate that the motor can deliver sufficient torque at this output it must adhere to 
\begin{equation}
    \sum_i P_{i,out,max}(\frac{v_{req}\cdot\gamma_i}{r_w}) \ge  F_x(v_{req})\cdot v_{req},
\end{equation}
where $P_{i,out,max}(\cdot)$ is the maximum available power evaluated at the given speed.

\paragraph{Acceleration Constraint $g_{pt,2}$}
The $0$–$100$ km/h acceleration time $t_{0-100}$ is evaluated through forward simulation and must satisfy
\begin{equation}
t_{0-100} \le t_{\max}.
\end{equation}
Where $t_{max}$ is the maximum admissible acceleration time.

\paragraph{Range Constraint $g_{pt,3}$}
\begin{equation}
\frac{E_{\mathrm{battery}}}{E_{\mathrm{in}}}
\cdot d_{\mathrm{WLTP}}
\ge R_{\min}.
\end{equation}
Where $E_{battery}$ is the battery capacity, $d_{WLTP}$ is the simulated driving distance and $R_{min}$ is the minimum range the vehicle should be able to travel.
\paragraph{Gradability Constraint $g_{pt,4}$}
\begin{equation}
T_{\mathrm{rear,max}}\cdot \gamma_{\mathrm{rear}}
+
T_{\mathrm{front,max}}\cdot \gamma_{\mathrm{front}}
\ge T_{\mathrm{grade,req}}.
\end{equation}
where $T_{\mathrm{grade,req}}$ is the torque at the wheel at a given gradability from which the vehicle should be able to start moving from.
% \subsubsection{Optimization Strategy}

% The objective is nonlinear and evaluated via time-domain simulation. Gradients are not available analytically due to interpolation-based efficiency maps and discrete feasibility checks. 
% A structured two-stage search strategy is therefore employed:
% Coarse grid exploration of $(s_{\mathrm{pmsm}}, s_{\mathrm{ac}})$ uteliziling analytical maxima and minima for $\gamma_{\mathrm{rear}}$ and $\gamma_{\mathrm{front}}$, and then a Local refinement around the best candidates. 
% This approach ensures feasibility through analytical gear ratio bounds, and provides deterministic subsystem responses suitable for hierarchical coordination.

\subsubsection{Optimization Strategy}

The objective is nonlinear and evaluated via time-domain simulation. Analytical gradients are unavailable due to interpolation-based efficiency maps and discrete feasibility checks.

A two-stage strategy is employed. First, coarse exploration is performed via grid-based sampling of $(s_{\text{pmsm}}, s_{\text{ac}})$, where for each sample $(\gamma_{\text{rear}}, \gamma_{\text{front}})$ are optimized jointly using the Nelder-Mead method \cite{Lagarias1998ConvergenceDimensions} within analytically derived bounds for the maxima and minima of of the gear ratios based on the performance constraints. Second, local refinement is performed around the best candidates.

% This formulation enforces feasibility through analytical bounds and yields subsystem responses suitable for hierarchical coordination.

\subsubsection{Subsystem Response}
Machine masses scale as
\begin{equation}
m_{pmsm} = m_{\mathrm{base,pmsm}}\cdot s_{pmsm} 
\end{equation}
\begin{equation}
m_{ac} = m_{\mathrm{base,ac}}\cdot s_{ac}
\end{equation}
and are returned to the system-level coordinator along with the net energy use given as cost function $J_{pt}$.

\subsection{Battery Pack Chassis Integration}

In the present work, the battery pack is modeled not only as an energy storage component but also as a structural element integrated within the vehicle floor. Its spatial placement influences both the global stiffness characteristics of the chassis and the overall mass distribution, thereby affecting the vehicle center of gravity (CoG) in both longitudinal and vertical directions. This model is implemented as an analytical model that purely evaluates the placement and returns its response in the form of a stiffness score used as an objective by the coordinator and a mass response. This means that the objective score of the model is fully dependent on the coordinator using SPI2 to explore the feasible design space.

% To enable efficient integration within the optimization framework, a reduced-order structural representation is adopted that captures the dominant contributions to bending and torsional stiffness while remaining computationally tractable.
\subsubsection{analytical model}
The vehicle underbody is represented as an equivalent beam section \cite{Fenner2012MechanicsStructures}, characterized by baseline bending and torsional stiffness values $K_{\text{bending,base}}$ and $K_{\text{torsion,base}}$. These correspond to a baseline second moment of area $I_{\text{base}}$ and torsional constant $J_{\text{base}}$.
% defined by the effective section width $b_{\text{eq}}$ approximated by the rocker spacing and baseline section height $h_0$. 
The integration of the battery pack modifies these section properties, yielding equivalent values $I_{\text{eq}}$ and $J_{\text{eq}}$. The resulting stiffness values are expressed through scaling relations
\begin{equation}
K_{\text{bending}} = K_{\text{bending,base}} \cdot \frac{I_{\text{eq}}}{I_{\text{base}}}, \end{equation}
\begin{equation}
K_{\text{torsion}} = K_{\text{torsion,base}} \cdot \frac{J_{\text{eq}}}{J_{\text{base}}},
\end{equation}
which follow directly from classical beam and torsion theory for structures with constant material properties \cite{Bauchau2009Euler-BernoulliTheory}.

The bending stiffness contribution of the battery pack is modeled through a parallel-axis formulation. The equivalent second moment of area is expressed as
\begin{equation}
I_{\text{eq}} = I_{\text{base}} + \eta_{\text{bend}} \cdot\, A_{\text{eff}}\cdot \, d^2,
\end{equation}
where $A_{\text{eff}}$ denotes the effective structural area of the battery pack, computed as a function of battery length $l_{\text{bat}}$ and effective width $b_{\text{eff}}$. The term $d$ represents the vertical distance between the battery centroid position $z_{\text{bat}}$ and the neutral axis of the equivalent section $z_{\text{NA}}$, such that $d = |z_{\text{bat}} - z_{\text{NA}}|$. The efficiency factor $\eta_{\text{bend}} \in [0,1]$ accounts for non-ideal load transfer and is defined as the product of physically motivated terms representing mounting stiffness, load path continuity, crossmember coupling, and longitudinal effectiveness. The latter is captured through a smooth position-dependent weighting function $\phi_{\text{bend}}(x_{\text{bat}})$ which accounts for the reduced structural influence of components located near the vehicle extremities., where $x_{\text{bat}}$ denotes the longitudinal battery position along the wheelbase $L$.

The torsional stiffness contribution is modelled through a combination of section closure and shear panel behaviour. The equivalent torsional constant is expressed as
\begin{equation}
J_{\text{eq}} = J_{\text{base}} + \eta_{torsion}\cdot\phi_{torsion}(x_{bat})\cdot \Delta J_{\text{panel}},
\end{equation}
where $\Delta J_{panel}$ represents an approximation of the torsion stiffness of a battery as a shear panel, $\eta_{torsion}\in [0,1]$ is used to define a load transfer efficiency $\phi_{torsion}(x_{bat})$ is a smooth weighting function for placement of the battery along wheelbase $L$ modelled after \cite{Young1976RoarksStrain}.

% where $\alpha_{\text{closure}}$ is a scaling parameter governing the contribution of section closure, $\phi_{\text{span}}$ is a geometric factor depending on the battery length $l_{\text{bat}}$, width $w_{\text{bat}}$, and their ratios with respect to the wheelbase $L$, and $\eta_{\text{torsion}} \in [0,1]$ represents the torsional load transfer efficiency. The additive term $\Delta J_{\text{panel}}$ captures the shear panel effect dependent on the estimated shear panel stiffness of battery integrated into the chassis construction as a torsion box

% In the adopted structural representation, the original crossmember layout is replaced by two structural supports located at the front and rear edges of the battery pack, denoted by positions $x_{\text{front}}$ and $x_{\text{rear}}$. These edge supports provide direct load paths between the battery pack and the rocker panels, forming the primary mechanism for load transfer in both bending and torsion. The structural effectiveness of these supports is represented through a factor $\phi_{\text{edge}}$, which depends on the presence of both supports and ensures a consistent torsional load path.

\begin{figure}[]
    \centering
    \includegraphics[width=\linewidth]{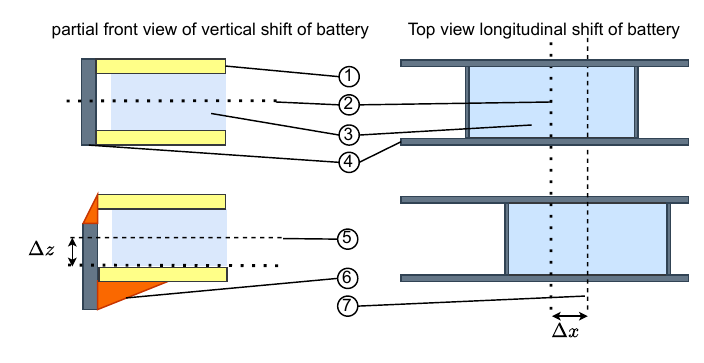}
    \caption{Depiction of battery integration in to chassis (not to scale): showing sandwich structure (1, yellow), Vertical and longitudinal neutral axis $z_{NA}$  and $x_{NA}$ (2, dotted), Battery (3, blue), Side member (4, Grey), Vertical shift of centroid position $z_{bat}$ (5, dashed), Mounting material (6, orange), longitudinal shift of centroid position $x_{bat}$ (7, dashed)}
    \label{fig:bat_in_chassis}
\end{figure}

% The structural contribution of the battery pack depends strongly on its spatial position. The longitudinal position $x_{\text{bat}}$ affects both bending and torsional stiffness through the weighting functions $\phi_{\text{bend}}(x_{\text{bat}})$ and $\phi_{\text{torsion}}(x_{\text{bat}})$, The vertical position $z_{\text{bat}}$ influences bending stiffness through the lever arm $d$ in the parallel-axis term.

In addition to stiffness contributions, the vertical placement of the battery pack introduces a trade-off in structural mass. The total vehicle mass is expressed as
\begin{equation}
m_{\text{underbody}} = m_{\text{base}} + m_{\text{battery}} + m_{\text{mount}}
\end{equation}
where $m_{\text{base}}$ denotes the baseline underbody mass, $m_{\text{battery}}$ is the battery mass, $m_{\text{mount}}$ represents additional mounting structure required to transfer loads between the battery and the side members. The mounting mass $m_{\text{mount}}$ is modelled as a linearly increasing function of the vertical battery position with the linear relation 
\begin{equation}
   m_{mount} =  \frac{1}{2}\cdot\Delta z\cdot(w_{sidemember}+w_{support})\cdot \rho_{steel}, 
\end{equation}
where $w_{sidemember}$ is the width of the side member, $w_{support}$ is the width of the lower support and $\rho_{steel}$ is the density of the steel, reflecting the need for additional structural material to bridge the gap between the battery and the primary load paths as presented by the orange mounting material that can be seen in Fig. \ref{fig:bat_in_chassis} which implementation is a simplified implementation of battery integration into the underbody with some crash structure consideration \cite{Belingardi2023BatterySolutions}.

Structural feasibility is enforced through constraints on bending and torsional stiffness,
\begin{equation}
K_{\text{bending}} \geq K_{\text{bending,target}}, 
\end{equation}
\begin{equation}
K_{\text{torsion}} \geq K_{\text{torsion,target}},
\end{equation}
where $K_{\text{bending,target}}$ and $K_{\text{torsion,target}}$ denote minimum required stiffness values.
% Violations are penalized through the addition of reinforcement mass $m_{\text{reinforcement}}$, which is modeled as proportional to the stiffness deficit.

% Within the overall ATC framework, the battery pack model provides a mapping from spatial placement variables $(x_{\text{bat}}, z_{\text{bat}})$ to structural stiffness, mass, and CoG location. Through its influence on both longitudinal and vertical CoG coordinates, the battery pack directly affects the coupling variable between placement and powertrain subsystems, enabling the optimization to exploit trade-offs between structural performance, mass distribution, and energy efficiency in a unified and physically consistent manner.
\subsubsection{intermediate discussion point}
For the purpose of design space exploration, this implementation is considered to be sufficient. The SPI2 approach already relies on approximations of component shapes for the placement, and thus for the exploratory design is the consideration is made to use an equivalent beam model to maintain computational tractability, and consider this highly simplified approach sufficient for the demonstration. As opposed to employing Finite Elememt Method or Body in White approaches which go into higher details. 

\subsubsection{subsystem response}
The battery-chassis integration subsystem response to the system level coordinator are the total underbody mass $m_{underbody}$ [kg] and a unitless stiffness score $K$ [-] based on the magnitude of the bending and torsion stiffness,
\begin{equation}
    K = K_{bending} +K_{torsion}.
\end{equation}

\subsection{Decomposition-Based Multi-Objective Optimization Using ATC-Inspired Coordination and NSGA-II}
\begin{figure}
    \centering
    \includegraphics[width=\linewidth]{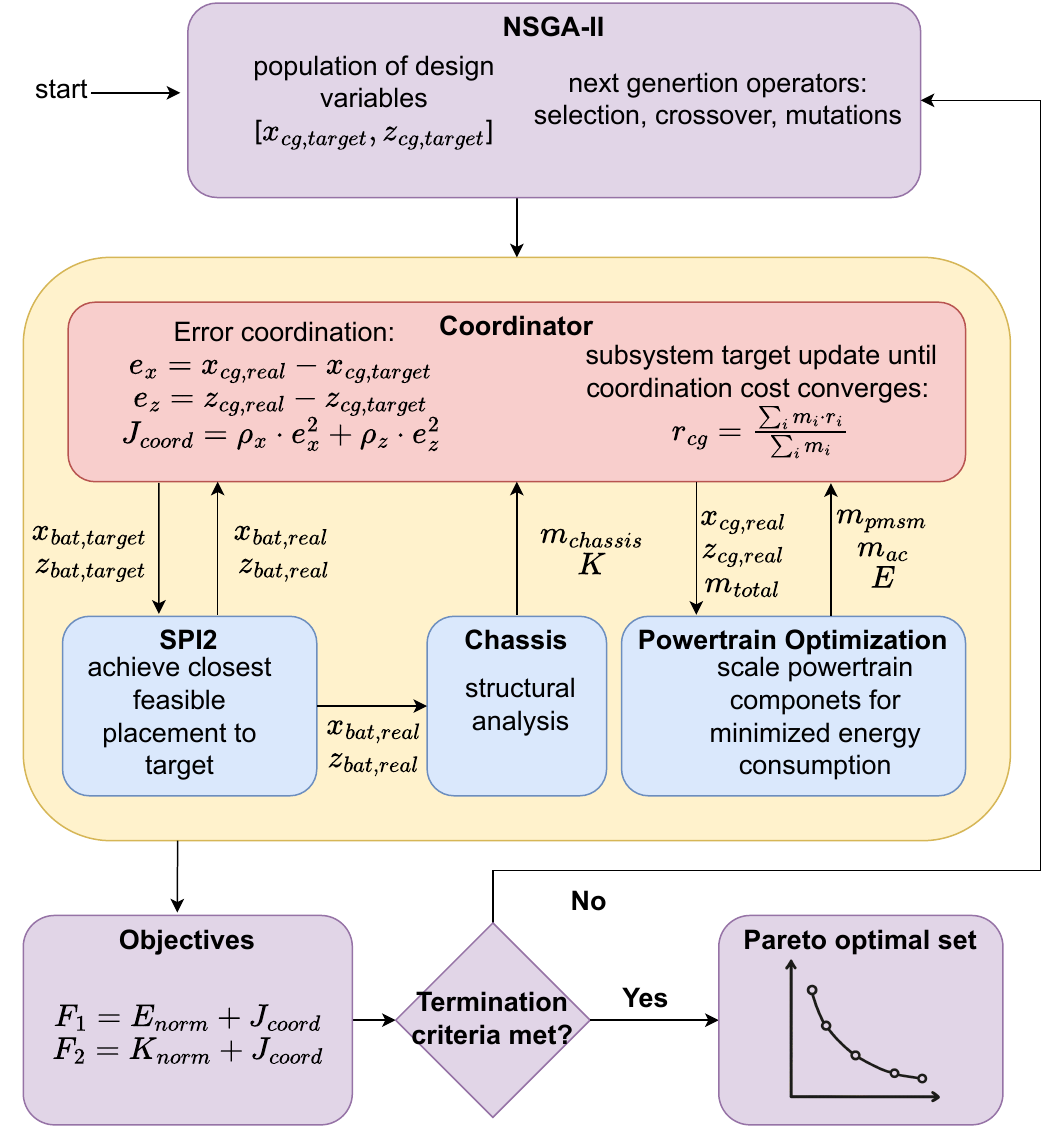}
    \caption{The ATC-inspired consistency coordination approach presented with incorporated NSGA-II as a multi-objective optimization handler in a schematic overview.}
    \label{fig:coordination}
\end{figure}
To jointly optimize battery placement, structural integration, and powertrain performance, a decomposition-based framework is employed in which subsystem interactions are coordinated ATC-inspired formulation \cite{Kim2003AnalyticalDesign} embedded within a multi-objective optimization process. This approach, as presented in Fig. \ref{fig:coordination}, enables the integration of heterogeneous subsystem models while preserving modularity, allowing each subsystem to be evaluated independently while still contributing to a consistent system-level design.

\subsubsection{Coupling Through Battery Placement}

The primary system-level design variables are the battery placement coordinates $x_{\mathrm{bat}}$ and $z_{\mathrm{bat}}$, which define the longitudinal and vertical position of the battery pack within the vehicle. The placement of the battery also induces a mass response, which contributes to the overall system mass and center of gravity $(x_{cg}, z_{cg})$. 

The system-level center of gravity is continuously recomputed as the mass-weighted sum of all subsystem contributions,
\begin{equation} \label{eq:cog}
r_{cg} = \frac{\sum_i m_i\cdot r_i}{\sum_i m_i},
\end{equation}
where $m_i$ denote the mass and $r_i$ defines both the $x$ and $z$ position of each contribution subsystem $i$. This formulation ensures that all spatial and parametric design changes are consistently propagated through the system. As a result, each new battery placement is evaluated based on the updated system CoG and mass distribution, reinforcing the bidirectional coupling between spatial design decisions and system-level performance.

The coupling between subsystems is therefore governed by the battery placement variables, with the center-of-gravity position acting as an intermediate physical quantity linking spatial design decisions to system-level performance.

\subsubsection{Target-Response Consistency}
The optimization process determines a target placement of the battery with corresponding target values $(x_{\mathrm{bat,target}}, z_{\mathrm{bat,target}})$ which are implicitly defined based on the system-level placement decision obtained through inversely using \ref{eq:cog} to determine the position of a component based on the target CoG. These targets are provided to the SPI2 subsystem and return the realized closest achievable feasible position $(x_{\mathrm{bat,real}}, z_{\mathrm{bat,real}})$.  The realized battery positioning is then used for the chassis subproblem, and the powertrain subproblem subsystem is supplied with the new system mass and updated system CoG obtained from Eq. \ref{eq:cog}.

The mismatch between target and realized system behavior is defined as
\begin{equation}
e_x = x_{\mathrm{cg,real}} - x_{\mathrm{cg,target}}, \qquad
e_z = z_{\mathrm{cg,real}} - z_{\mathrm{cg,target}}.
\end{equation}

These residuals quantify the degree to which the intended system-level response is achieved by the subsystem evaluations and serve as measures of consistency within the decomposed framework.

\subsubsection{ATC-Inspired Coordination Using Quadratic Penalty}
Consistency between subsystem targets and responses is enforced using a quadratic coordination method, which represents a simplified form of ATC coordination. Due to the coordination through the Non-dominated Sorting Genetic Algorithm II  (NSGA-II) as described in the following part, there is no information on the history of iterations. So instead of employing a full augmented Lagrangian formulation with iterative multiplier updates, the present approach incorporates consistency directly into the objective functions through a quadratic penalty term:
\begin{equation}
J_{\mathrm{coord}} = \rho_x \cdot e_x^2 + \rho_z \cdot e_z^2,
\end{equation}
where $\rho_x$ and $\rho_z$ are penalty parameters controlling the strength of coordination in the longitudinal and vertical directions, respectively.

This quadratic coordination method penalizes deviations between target and realized system responses, encouraging solutions that satisfy subsystem consistency while maintaining a continuous and well-scaled optimization landscape. 
% To allow minimal mismatch of the placement, relaxed bounds are imposed:
% \begin{equation}
% |e_x| \le \varepsilon_x, \qquad |e_z| \le \varepsilon_z.
% \end{equation}

% These bounds act as relaxation for the , while the primary coordination mechanism remains the quadratic penalty.

\subsubsection{Multi-Objective Formulation}

The integrated design problem is formulated as a multi-objective optimization problem that simultaneously considers energy efficiency and structural performance. The objectives are defined as
\begin{equation}
F_1 = E_{\mathrm{norm}} + J_{\mathrm{coord}},
\end{equation}
\begin{equation}
F_2 = K_{\mathrm{norm}} + J_{\mathrm{coord}},
\end{equation}
where $E_{\mathrm{norm}}$ represents the normalized energy consumption over the driving cycle and $K_{\mathrm{norm}}$ represents the normalized structural stiffness metric.

By augmenting both objectives with the coordination penalty, the optimization process evaluates solutions not only based on performance metrics but also on the consistency between subsystem responses. This ensures that high-performing solutions are physically coherent across subsystems.

\subsubsection{NSGA-II Optimization}

The resulting multi-objective optimization problem is solved using the NSGA-II \cite{Deb2002ANSGA-II}. Within this framework, NSGA-II operates as the system-level coordination mechanism by iteratively sampling candidate battery placements and evaluating their corresponding subsystem responses.

% The algorithm maintains a population of design candidates, which are evaluated based on the augmented objective functions. Through non-dominated sorting, solutions are ranked according to Pareto dominance, while crowding distance is used to preserve diversity within the population. This enables the algorithm to approximate the Pareto-optimal front representing the trade-off between energy consumption, structural stiffness, and subsystem consistency.

Unlike classical ATC implementations that rely on sequential coordination and gradient-based updates, NSGA-II does not require derivative information and is well suited to problems involving nonlinear, discontinuous, or simulation-based subsystem evaluations. Its population-based nature allows simultaneous exploration of multiple regions in the design space, improving robustness and reducing the risk of convergence to local optima.

In this context, NSGA-II effectively replaces the traditional ATC coordination loop by embedding the consistency enforcement directly within the objective evaluation, enabling a unified and efficient exploration of coupled design variables and system-level performance trade-offs.

%% file: chapters/results.tex
This section covers the results. First, the SPI2 findings are presented.
Then, the integration for vehicle designs is analysed using models to demonstrate the applicability of the SPI2 model in larger-scale optimization. \\
\textit{Hardware used to evaluate computational load:} AMD Ryzen 7 pro 8845HS CPU, NVIDIA RTX 1000 (6GB) GPU, 32 GB RAM running Python version 3.11.9 and CasADi 3.7.2.

\subsection{SPI2 integrability}
\begin{figure}[]
    \centering
    \includegraphics[width=\linewidth]{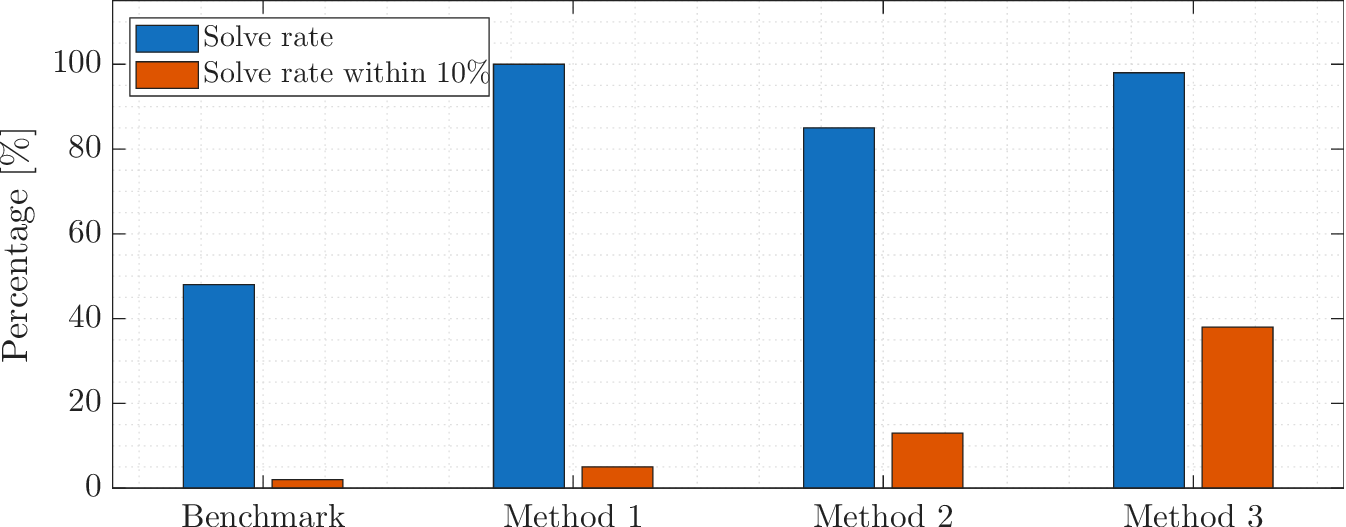}
    \caption{Percentage of boundary box constraint placement problem initializations converging to a solution and percentage of solutions within 10\% of best solution}
    \label{fig:convergence_quality}
\end{figure}

To make the SPI2 implementable for larger-scale optimization problems, the computational load must be reduced, and the quality of the solver must be increased to more reliably obtain solutions of higher quality. Therefore, the effects of the presented methods are shown benchmarked against methods described by existing literature. Table \ref{tab:placement_methods} present the benchmark approach and the changes in the rotational approach and the boundary box constraint approach for each method. The problem used for evaluation against the benchmark was a volumetric minimization of 6-cuboid object of 1 by 1 by 0.5 units described by 14 spheres each, constrained within a flat boundary box of 15 by 15 by 1.5 units. The SDF approach was constructed on a grid with a resolution of 128$^3$, and the boundary box sphere approach was constructed using 200 spheres on a 128$^3$ grid method is tested for the same set of 100 randomized initial positions. In Fig. \ref{fig:convergence_quality}, the convergence performance of each method is presented, showing that an improvement in solve rate can be obtained both through using Quaternion rotation or the SDF approach, as opposed to Euler rotation and boundary box spheres. Additionally, a significant increase in the quality of the solutions can be observed from the percentage of solutions being within 10\% of the best found solution.

Evaluating the computational loads between the methods as presented in Tab. \ref{tab:improvements_against_benchmark}. Firstly, it can be observed that the use of quaternion rotation reduces the mean number of iterations required to reach convergence. This may be explained by Euler rotation ineffectively rotating due to an instance of gimbal lock, which also explains the lower solve rate for the benchmark method and method 2. Secondly, the use of an SDF reduces the mean computation time per iteration as the computational load is reduced. Each sphere $\mathcal{S}$ now only requires a single interpolation of the SFD as opposed to an evaluation of its position to all boundary box spheres. The lower volume obtained by method 3 can be explained by the object not needing to rely on the coverage of boundary box spheres to be placed close together through the use of an SDF, the alignment of the objects within an axis aligned bounding box causes the rotational axis's of the objects to align with the workspace axis which induces gimbal lock for the Euler rotation. 
With all these considerations, Method 3 may be assumed to be the best option for the desired implementation. A visualization of the best obtained result of each of the methods can be seen in Appendix \ref{appendix:Appendix_placement_methods}.

\begin{table}[]
\caption{Set of SPI2 placement strategies to quantify improvements against benchmark strategy}
\label{tab:placement_methods}
\begin{tabular}{l|l|l}
Method & Rotational Approach & Boundary box constraint Approach\\ \hline
Benchmark          & Euler             & Boundary Box Spheres          \\
Method 1           & Quaternion        & Boundary Box Spheres          \\
Method 2           & Euler             & Signed Distance Field         \\
Method 3           & Quaternion        & Signed Distance Field        
\end{tabular}
\end{table}

\begin{table}[]
\caption{Computational load and performance comparison of proposed methods against the benchmark.}
\label{tab:improvements_against_benchmark}
\begin{tabular}{l|l|l|l}
 & Mean no. iterations & Time per iteration [s] & Best solution \\ \hline
Benchmark          & 1072   & 0.4560    & 2.559                \\
Method 1           & 301    & 0.4691    & 2.494                \\
Method 2           & 2607   & 0.1259    & 2.596                \\
Method 3           & 257    & 0.1453    & 1.728               
\end{tabular}
\end{table}

\subsection{SPI2 usecase results}
\begin{figure}[]
    \centering

    % Row 1: single image
    \begin{subfigure}{\linewidth}
        \centering
        \includegraphics[width=\linewidth]{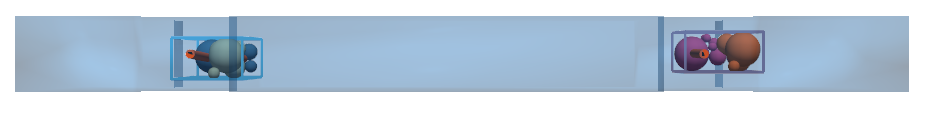}
        \caption{side-view}
    \end{subfigure}

    \vspace{0.5cm}

    % Row 2: single image
    \begin{subfigure}{\linewidth}
        \centering
        \includegraphics[width=\linewidth]{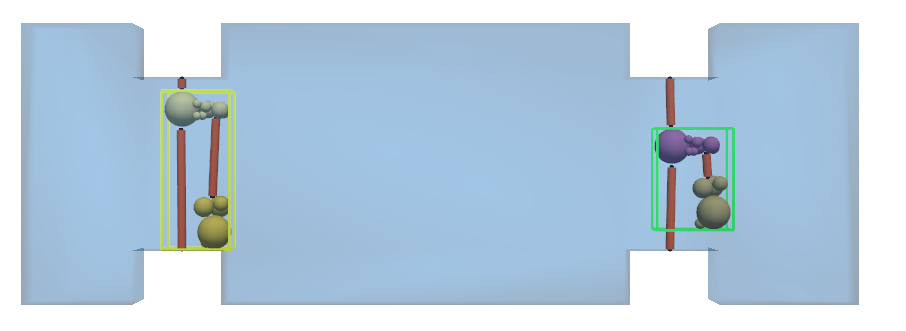}
        \caption{top-view}
    \end{subfigure}

    \vspace{0.5cm}

    % Row 3: two images side by side
    \begin{subfigure}{0.49\linewidth}
        \centering
        \includegraphics[width=\linewidth]{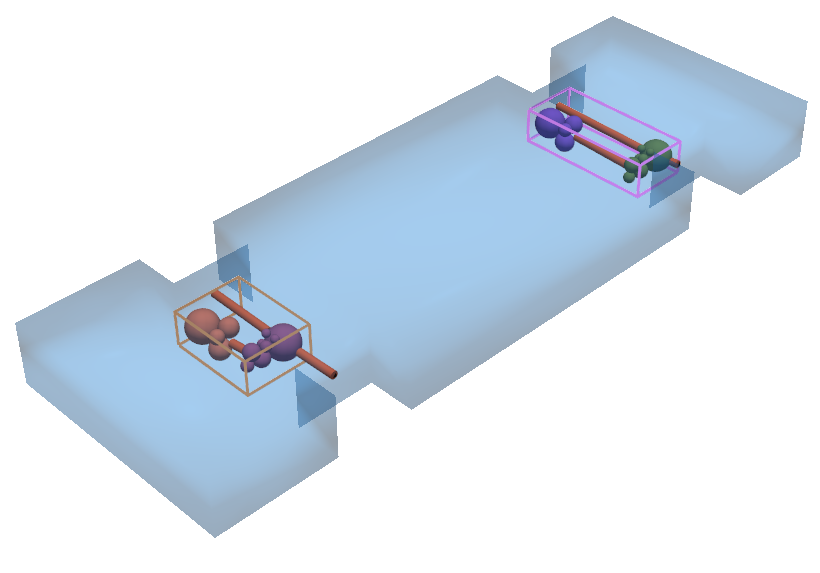}
        \caption{Iso-view}
    \end{subfigure}
    \hfill
    \begin{subfigure}{0.49\linewidth}
        \centering
        \includegraphics[width=\linewidth]{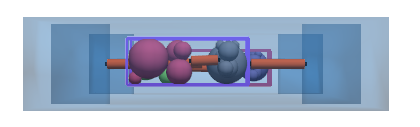}
        \caption{front-view}
    \end{subfigure}

    \caption{Result for the placement problem to place powertrain components in the skateboard model. The side, top and front-view show that all components are inside the feasible region, and that the mechanical axles are aligned within the specified accepted range of the desired direction.}
    \label{fig:placement_results}
\end{figure}

Utilizing the improvement in the SPI2 framework, it may now be employed to the vehicle use case. A skateboard model of a vehicle base is given as a boundary box, the front and rear axle motors and differential gearboxes are introduced as MDBD objects and the set of connections defining the interconnected system are provided. The SPI2 framework generates the optimized powertrain layout adhering to all the constraints which can eb seen in Fig. \ref{fig:placement_results}. This layout is then used to continue on with the decomposed placement problem to optimize the battery placement. A limitation in the current framework can already be observed from the top-view in Fig. \ref{fig:placement_results} that the competing objectives of alignment and spatial minimization lead to local optima and not global optima, as seen in the front axle powertrain converging to a different positioning compared to the rear axle.

% \begin{figure}
%     \centering
%     \includegraphics[width=\linewidth]{Figures/solver_powertrain.eps}
%     \caption{Objective and constraint violation values during optimization}
%     \label{fig:solver_powertrain}
% \end{figure}

\subsection{Battery-chassis integration results and powertrain results}
% Hier misschien verwijzen naar stiffness gain gevonden in body in white paper vergelijkbare improvments

\begin{figure}[]
    \centering
    \begin{subfigure}{0.49\linewidth}
        \centering
        \includegraphics[width=\linewidth]{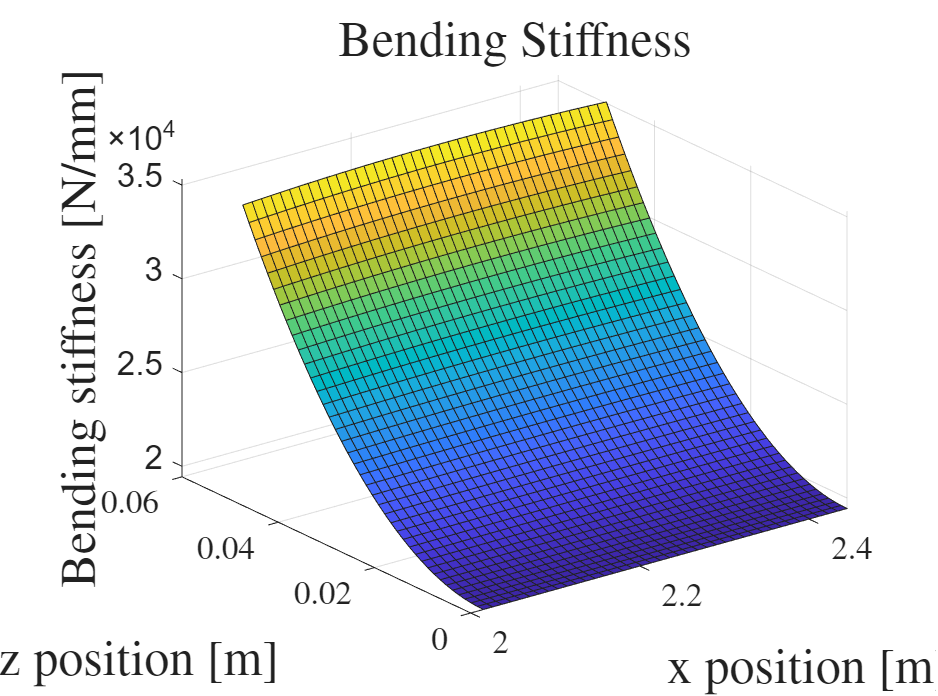}
        \caption{Bending stiffness}
    \end{subfigure}
    \hfill
    \begin{subfigure}{0.49\linewidth}
        \centering
        \includegraphics[width=\linewidth]{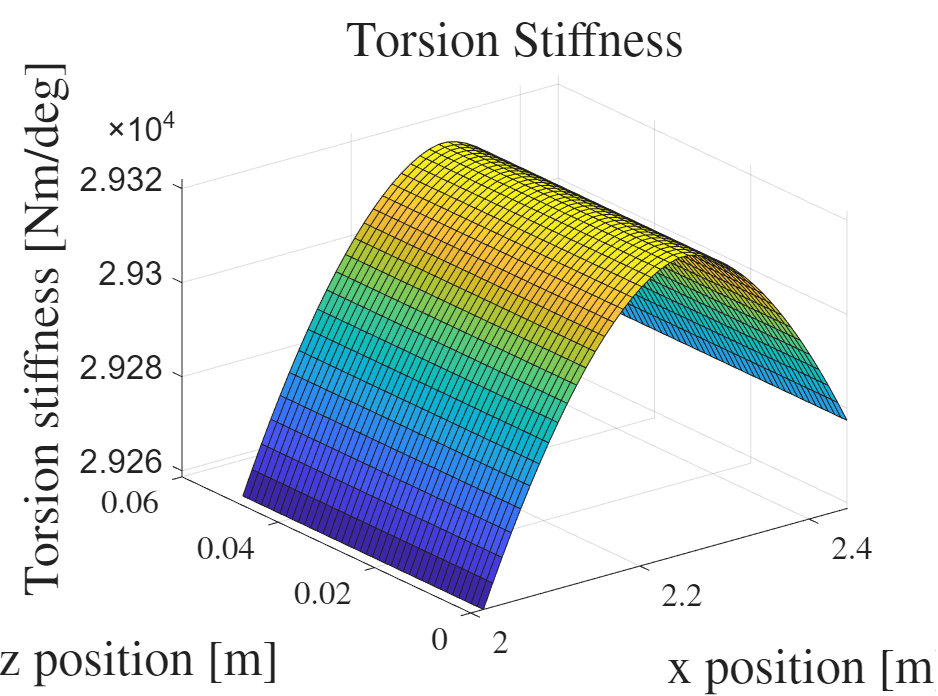}
        \caption{Torsion stiffness}
    \end{subfigure}

    \caption{Equivalent section beam model approximations on changes in stiffness based on location-dependent integration of battery pack into chassis under the assumption that placement is feasible within this range}
    \label{fig:stiffness_plots}
\end{figure}

\begin{figure}[]
    \centering
    \begin{subfigure}{0.49\linewidth}
        \centering
        \includegraphics[width=\linewidth]{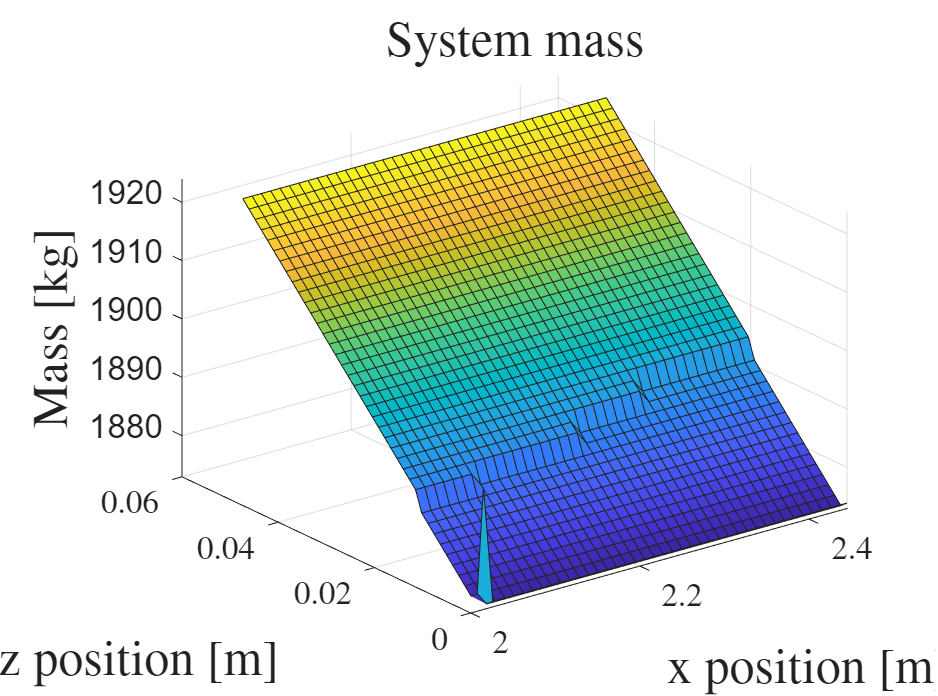}
        \caption{total system mass}
    \end{subfigure}
    \hfill
    \begin{subfigure}{0.49\linewidth}
        \centering
        \includegraphics[width=\linewidth]{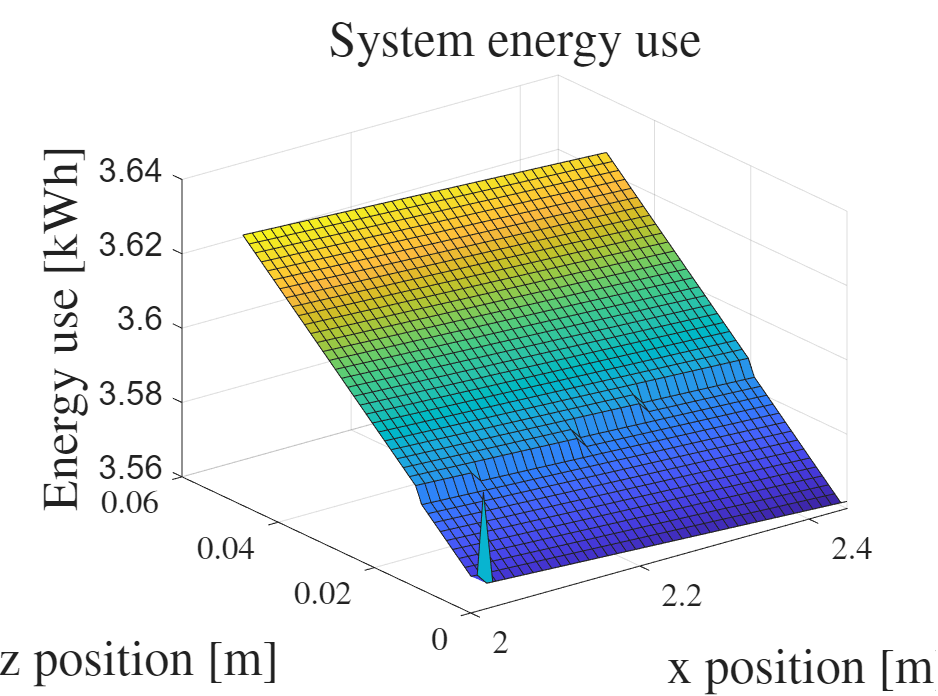}
        \caption{total energy use over a WLTP cycle}
    \end{subfigure}

    \caption{combined obtained from combined subsystems and system energy use over a WLTP cycle based on location-dependent integration of battery pack into chassis under the assumption that placement is feasible within this range}
    \label{fig:system_mass_energy}
\end{figure}

The resulting impact of the integration of the battery pack into the chassis and the impact of the longitudinal and vertical centroid position of the battery pack can be seen in Fig. \ref{fig:stiffness_plots}. The base values assumptions used for bending stiffness $K_{base,bending}$ were 18000 [N/mm] and $K_{base,torsion}$ is taken at 22000 [Nm/deg] based on \cite{Danielsson2016InfluenceCharacteristics} and the obtained stiffness values fall within an acceptable of the battery pack as structural element contributing an average of 44\% additional bending stiffness and 32\% torsional stiffness, the battery structure mass reaching reaching up to around 60 [kg]  is also reasonable \cite{Rosso2021EfficientVehicles}. Considering the low fidelity of the equivalent section model intended to explore design trade-offs, this quality is considered sufficient for the use case.
The system mass and system energy use presented in Fig. \ref{fig:system_mass_energy} present the strong correlation between the between weight savings obtained through the integration of the battery in the vehicle chassis, and the system energy use over the WLTP cycle. another slight correlation can be observed from the light inconsistency in the otherwise smooth plane where the optimal region of the powertrain moves to a different region due to the non-convexity of the efficiency maps and the motor masses change. lastly the system energy has a slightly downward slope whit the increase of the X coordinate of the battery moves to the rear of the vehicle as it increases the load on the rear axle causing the a change in the brake bias to increase the regenerative braking fraction on the rear axle which has more favourable efficiency map for regenerative braking.

% \textcolor{red}{surf plots mass and energy use nog bij zetten}

\subsection{Multi-objective optimization results}
\subsubsection{Numerical results}
\begin{figure}
    \centering
    \includegraphics[width=\linewidth]{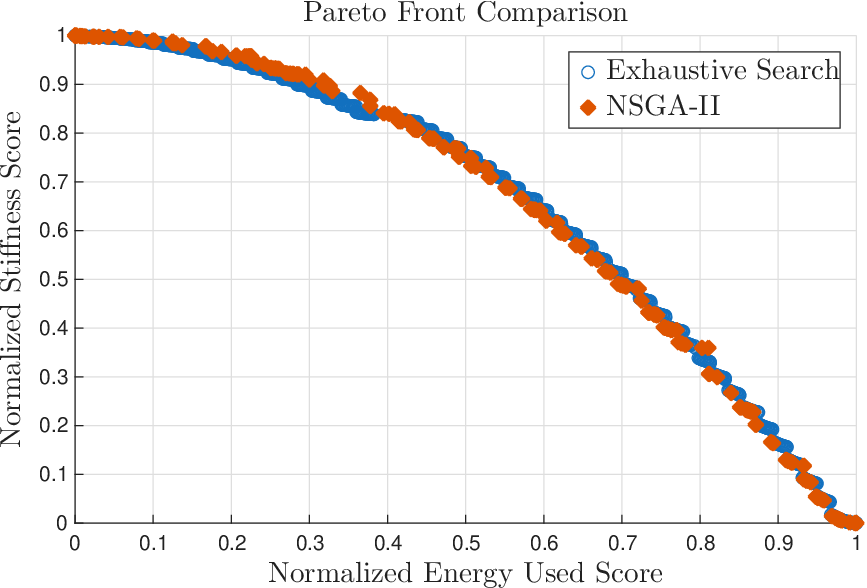}
    \caption{Pareto front obtained through Brute force approximation through gridded search and NSGA-II optimization of normalized energy use against normalized stiffness}
    \label{fig:pareto_front}
\end{figure}

\begin{figure}
    \centering
    \includegraphics[width=\linewidth]{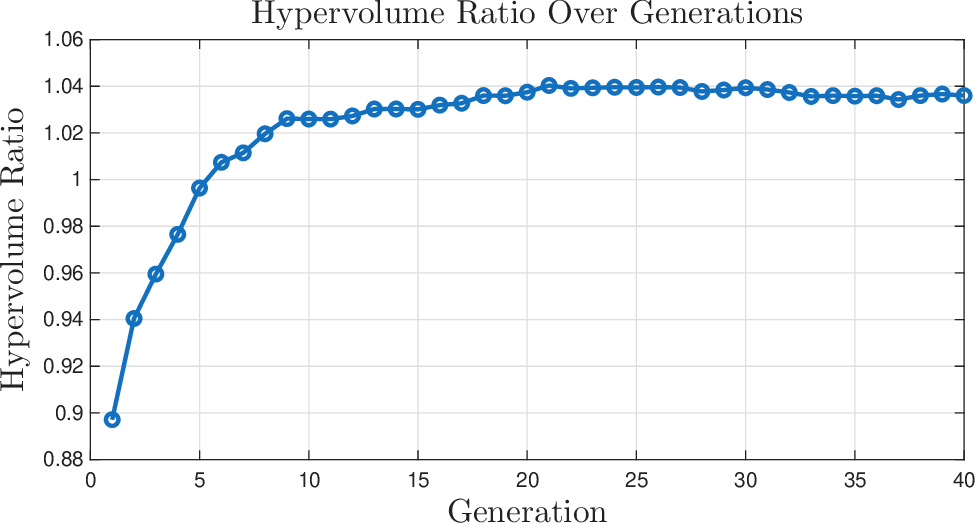}
    \caption{Pareto front hypervolume ratio per generation between NSGA-II and Brute force approach}
    \label{fig:hyper volume}
\end{figure}

As the real objective value ranges can be observed in the previous results a continuation is made with normalized result values in the range $[0,1]$ where 0 is the best obtainable solution in the evaluated set and 1 is the worst result in the set. To quantify the quality of the SPI2 integration with NSGA-II a comparison was made to a brute force approximation of the Pareto front through an exhaustive search over a discretized grid over the feasible range of the battery pack location with 5mm accuracy for the x-coordinate and 1mm accuracy for the z-coordinate. The Pareto front results can be observed in Fig \ref{fig:pareto_front}. The benefit of the combined use of NSGA-II as an optimization coordinator with SPI2 as a sub problem to enforce feasibility can be observed here as it enhances the applicability of the SPI2 as its placement target coordinates can be fully continuous. To this extent the improvement in finding the Pareto front is presented through the hypervolume (HV) ratio which is explained in Appendix \ref{sec:Pareto_metricx} and resutlts are presented in Fig. \ref{fig:hyper volume}, where it can be seen that for the given exhaustive search discretization the NSGA-II approach obtains an improved result after 6 generations with each generation existing of 40 evaluations at which point the HV ratio exceeds 1 and converges to 1.036. 
An additional note is the the first improvement at the 6th generation indicates that an improved results is obtainable in only 240 evaluations, opposed to 4080 evalutions required for the exhuastive search indicating and factor 17 decrease in required evaluations, which once again demonstrates the computational tractability of the use of combined NSGA-II with SPI2 in optimization. 

% \begin{figure}
%     \centering
%     \includegraphics[width=\linewidth]{Figures/Pareto_front.eps}
%     \caption{Pareto front of normalized energy density of the integrated battery pack in the chassis and the normalized chassis stiffness score}
%     \label{fig:pareto_front}
% \end{figure}
% 

Some furhter results that show the accuracy of the SPI2 in the decomposed optimization are presented in Tab. \ref{tab:pareto_performance}, highlighting the low Inverted Generational Distance (IGD) used as a measure defining the coverage of the Pareto front. Additionally, the maximum and average placement error are presented, as the error in placement is utilized in the quadratic penalty of the coordination strategy.  It is important that the objective and coordination penalty lead to physically feasible results and do not largely exceed the target placement, as the maximum error is within 2 [mm] of the target and the average error under 0.1 [mm] it is considered to lead to sufficiently accurate results for the automotive use case. For other use cases the scale of the error could always be furhter tuned through the quadratic penalty terms
\begin{table}[]
\caption{evaluation metrics of combined SPI2 and NSGA-II performance in optimization}
\label{tab:pareto_performance}
\begin{tabular}{l|l|p{0.3\linewidth}|p{0.3\linewidth}}
HV ratio & IGD   & Maximum placement error {[}mm{]} & Average placement error {[}mm{]} \\ \hline
1.036    & 0.025 & 1.6                              & 0.08                           
\end{tabular}
\end{table}

\subsubsection{Computation time}
To evaluate the computational tractability of the SPI2 integration into evaluation of design generation a wall clock time metric was taken to evaulate the computation time of all sub problems. the comparison of computational time is evaluated between the exhaustive search approach and the 6th generation of the NSGA-II as the 6th generation was the first generation to obtain a comparable, or better yet, improved result to the brute force based on the HV ratio. The exhaustive search approach consisted of a runtime of 136 hours, the 6th generation NSGA-II was obtained at a runtime of 6.6 hours meaning a $95.11\%$ reduction in compution time is achieved through utelizing placement feasibility as a coordination approach as opposed to brute forcing each possible placement and running a separate evaluation to validate the feasibility of the position. In both the exhaustive search and NSGA-II the powertrain optimization accounted for about $20\%$ of the runtime, SPI2 accounted for about $80\%$ of the runtime and the chassis evaluation was $<1\%$.

% - pareto front 

% -NSGA-II approach

% -ATC approach

%% file: chapters/discussion.tex
The improvement of the SPI2 is presented using a benchmark problem highlighting improvements that can be obtained. As the quality of the solution is highly dependent on the complexity of the problem, no hard statements can be made on the improvements, and only improvement trends can be observed.

One goal of the presented research was to find a meaningful integration of SPI2 into system-level optimization as an effective way of maintaining design feasibility of component placement during optimization. with SPI2 integrated into the framework, with the intention of an early-stage design space exploration tool. As such, the employability of the SPI2 in system-level design as a coordination strategy enforcing feasibility was investigated as presented in the paper. This implementation was performed around a vehicle use case loosely modelled around Skoda Enyaq to obtain a reasonable baseline of the vehicle; all further evaluations were performed around benchmark results of the modelled vehicle.

The full framework setup with the chosen sub-problems to be integrated is based on its applicability to utilize the SPI2. Therefore, no statements are provided on the quality of the powertrain optimization and the chassis stiffness evaluation. These evaluations are based on simplified physics models and were chosen as they are both aspects that are influenced by the placement of the battery and its effect on mass distribution. As well as their effect on the total system mass, through the machine scaling in the powertrain optimization, and the mass of the required mounting material for the battery chassis integration. These sub-problems are demonstrative problems intended to showcase the system-level capabilities of exploring the design space effectively.

%% file: chapters/conclusion.tex
This paper introduced adaptations to previous approaches for SPI2 frameworks such that it obtains improved results more reliably at a lower computational cost.  
The SPI2 framework was extended to be used as a generative design tool capable of generating the physical layout and alignment of powertrain components and their mechanical axles. 
Furthermore, the use of an SPI2 framework was investigated to be employed for higher-level system design intended for exploring design trade-offs between varying design goals using a decomposed optimization approach. The result show that the utilization of SPI2 to evaluate placement feasibility and utilizing the placement error to formulate a coordination strategy can be successfully implemented. Besides successful implementation it is shown to improve the findings as presented by the increased HV ratio value, which can be achieved at a lower computational cost with a runtime reduction of 95.11\% opposed to brute force. At a placement accuracy within 2 millimeters.

%% file: chapters/future_work.tex
Future work can be focused on several directions. First, it may focus on extending the geometric modelling beyond the current MDBD representation toward higher-fidelity object approximations, capturing more details, enabling applicability beyond early-stage design. 
Second, in the presented work a proof of concept was presented for the integration of SPI2 to include physical positioning as a meaningful, constrainable and enforceable variable to ensure design feasibility. The usability of this finding could be further investigated for other uses besides powertrain such as payload packaging to improve mass distribution.
Lastly, other types of problem decomposition approaches may be investigated and benchmarked to determine is there a better ways of integrating SPI2 systems.

%% file: chapters/acknowledgement.tex
%The author would like to thank Prof. Dr. Ir. Theo Hofman and Ir. Jorn van Kampen for their guidance and supervision of the research project.
The author acknowledges that during the development of this work, generative AI tools (OpenAI ChatGPT) have been used. This use was employed to improve clarity and structure. All analysis and technical content is the author's original work.

%% file: chapters/appendix.tex
\section{SPI2 approach methods results}
\label{appendix:Appendix_placement_methods}
\begin{figure}[H]
    \centering
    \includegraphics[width=\linewidth]{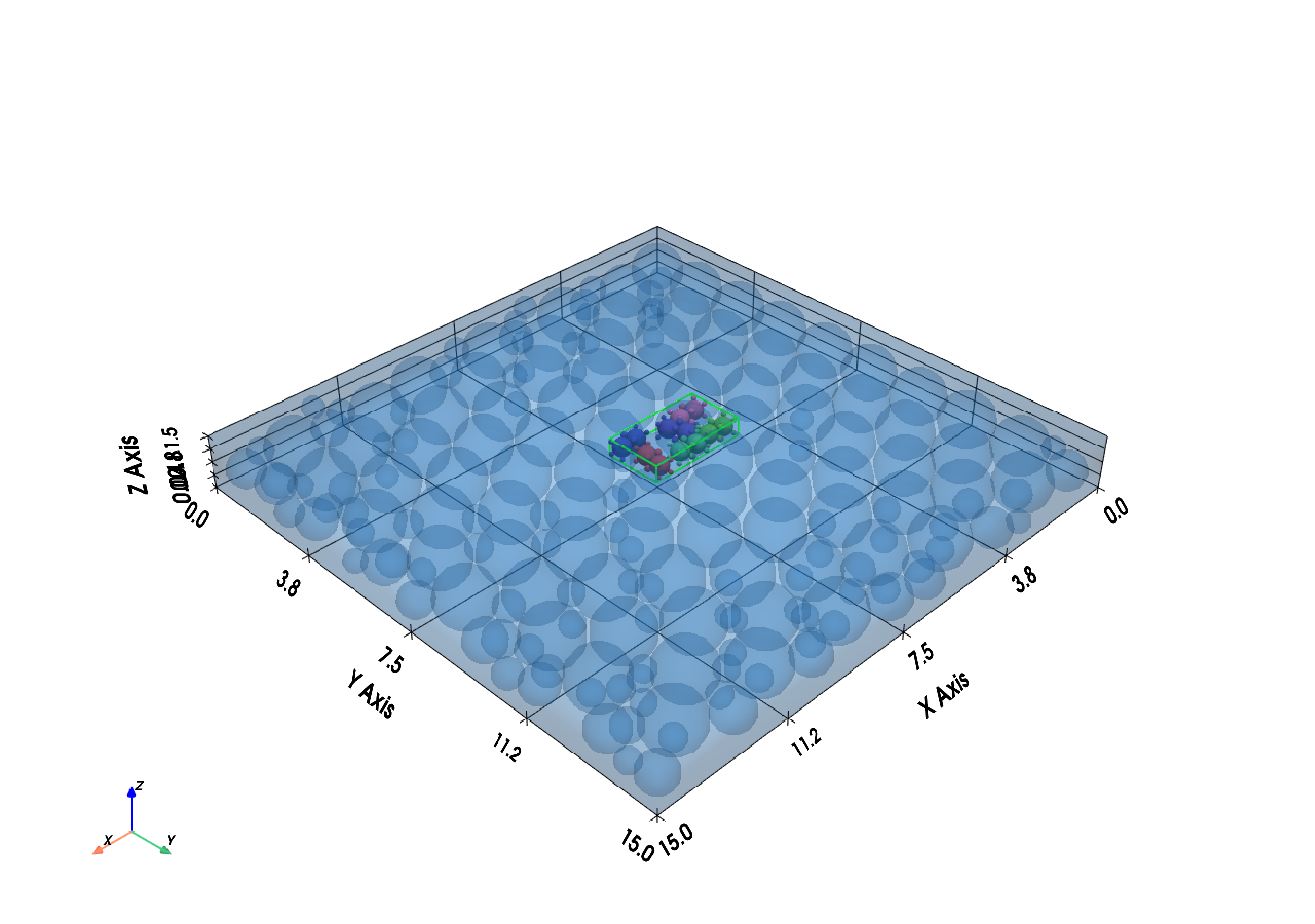}
    \caption{Best obtained result for the Benchmark Euler rotation and Boundary Box Sphere approach}
    \label{fig:benchmark_result}
\end{figure}

\begin{figure}[H]
    \centering
    \includegraphics[width=\linewidth]{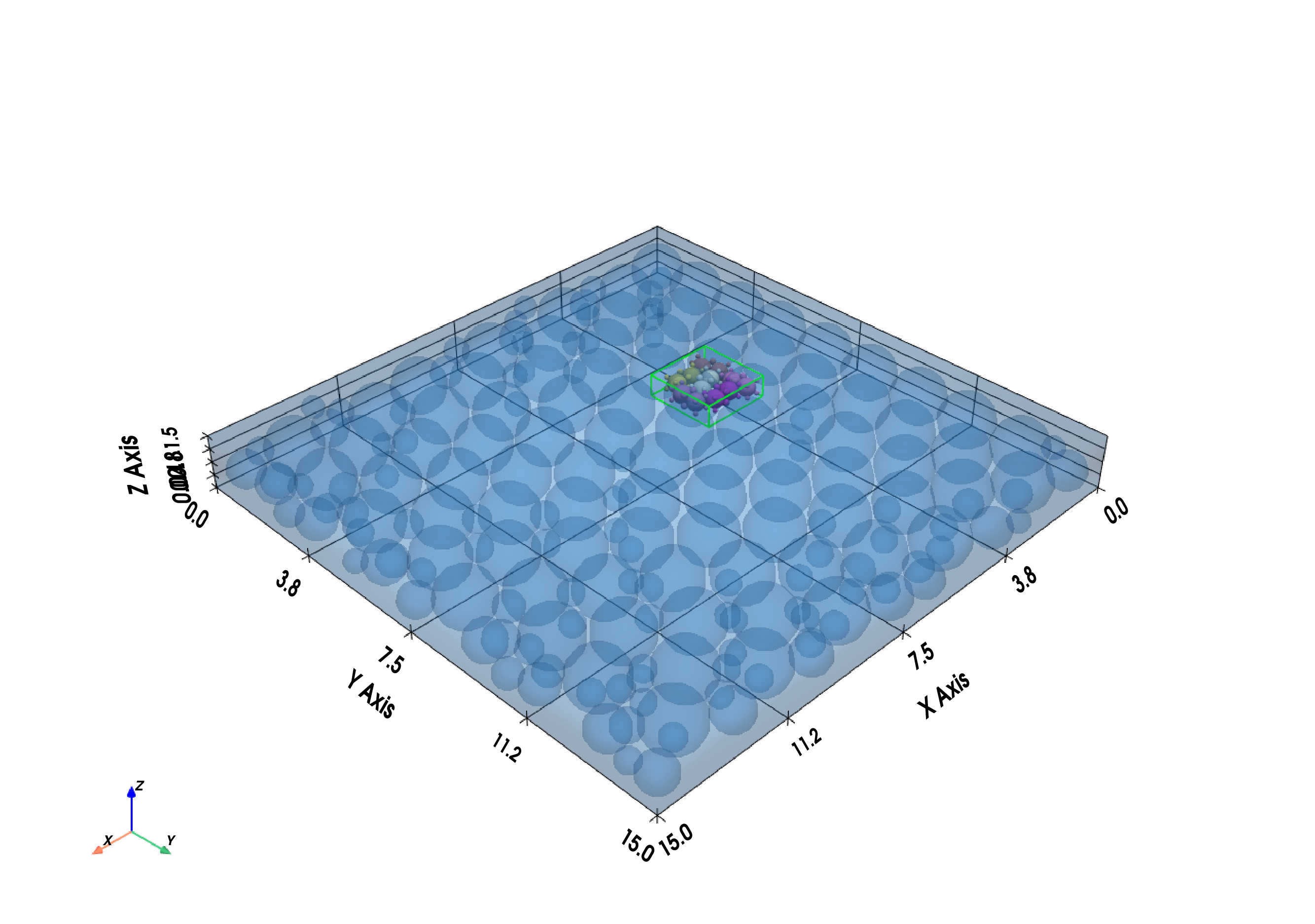}
    \caption{Best obtained result for method 1 Quaternion rotation and Boundary Box Sphere approach}
    \label{fig:method2_result}
\end{figure}

\begin{figure}[H]
    \centering
    \includegraphics[width=\linewidth]{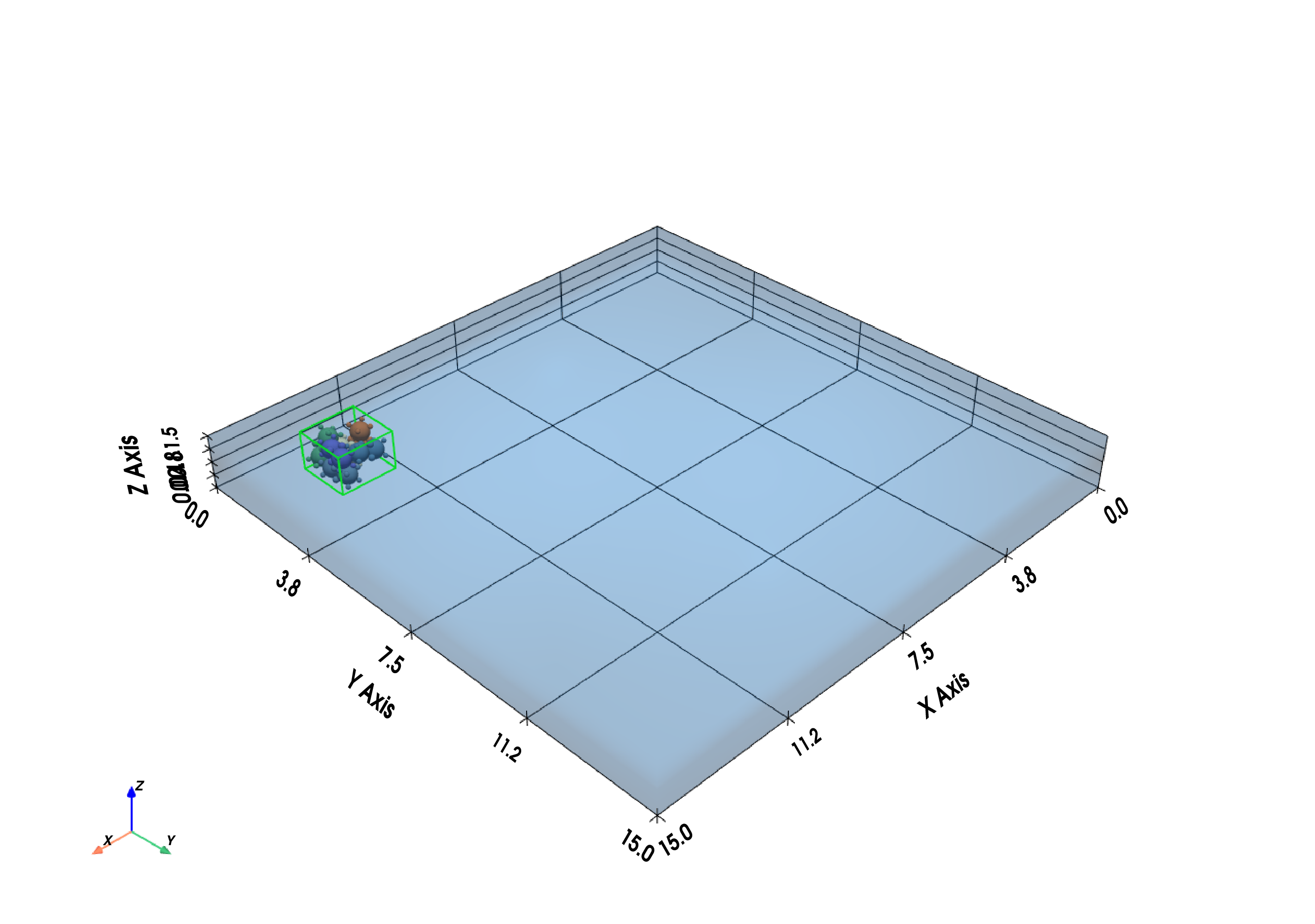}
    \caption{Best obtained result for method 2 Euler rotation and Signed Distance Field approach}
    \label{fig:method3_result}
\end{figure}

\begin{figure}[H]
    \centering
    \includegraphics[width=\linewidth]{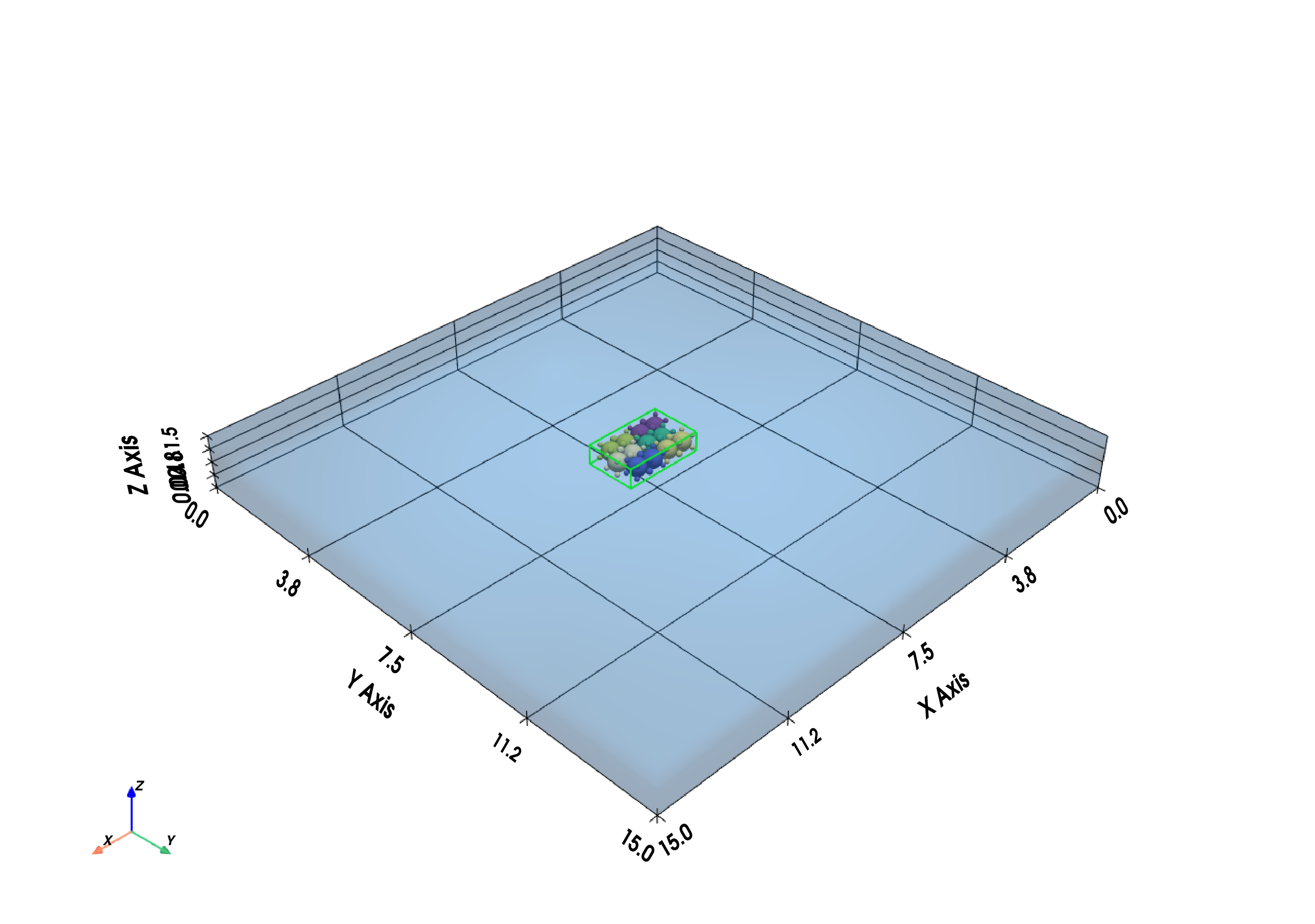}
    \caption{Best obtained result for method 3 Quaternion rotation and Signed Distance Field approach}
    \label{fig:method4_result}
\end{figure}

\section{evalutuation metric Pareto front}\label{sec:Pareto_metricx}
\subsection{Hypervolume (HV)}
The Hypervolume (HV) metric evaluates the quality of a Pareto front by measuring the objective-space area dominated by the obtained non-dominated solutions relative to a predefined reference point \cite{Audet2021PerformanceOptimization}. For a minimization problem, a larger hypervolume indicates a better Pareto front, since it represents solutions that are closer to the ideal objective values and cover a wider portion of the trade-off space.

For the two-objective case used in this work, the Pareto front consists of points:

\begin{equation}
F = \left\{ \left(f_1^i, f_2^i\right) \right\}_{i=1}^{N}
\end{equation}

where:

\begin{itemize}
    \item $f_1$ = normalized energy objective
    \item $f_2$ = normalized stiffness objective
\end{itemize}

A reference point is defined as:

\begin{equation}
\mathbf{r} = (r_1, r_2)
\end{equation}

which must be worse than all Pareto points in both objectives. In this work, the normalized reference point was chosen as:

\begin{equation}
\mathbf{r} = (1.1,\;1.1)
\end{equation}

The Pareto points are first sorted in ascending order of $f_1$. Dominated points are removed, after which the hypervolume is computed as the sum of rectangular areas between consecutive Pareto points and the reference point:

\begin{equation}
HV = \sum_{i=1}^{N}
(r_1 - f_1^i)\,(f_2^{i-1} - f_2^i)
\end{equation}

with:

\begin{equation}
f_2^{0} = r_2
\end{equation}

Only rectangles with positive width and height contribute to the hypervolume.

\subsection{Hypervolume Ratio (HV Ratio)}

To compare the NSGA-II Pareto front against the brute-force reference Pareto front, the Hypervolume Ratio was used. This metric expresses how much of the reference Pareto front hypervolume is captured by the approximated front or exceeding it \cite{Audet2021PerformanceOptimization}. 

The hypervolume ratio is defined as:

\begin{equation}
HV_{\text{ratio}} =
\frac{HV_{\text{approx}}}
{HV_{\text{reference}}}
\end{equation}

where:

\begin{itemize}
    \item $HV_{\text{approx}}$ = hypervolume of the NSGA-II Pareto front
    \item $HV_{\text{reference}}$ = hypervolume of the brute-force reference Pareto front
\end{itemize}

Both fronts are normalized using the minimum and maximum objective values of the reference Pareto front before calculating the hypervolume. 

%% file: references.bib
@article{Chen2022,
    title = {{A Topology Optimization Process for Discrete Modular Design based on Discrete Element Modelling for generating reconfigurable funicular structures}},
    year = {2022},
    author = {Chen, Qinglu}
}

@article{VanKampen2023,
    title = {{A Two-dimensional Spatial Optimization Framework for Vehicle Powertrain Systems}},
    year = {2023},
    journal = {2023 IEEE Vehicle Power and Propulsion Conference, VPPC 2023 - Proceedings},
    author = {Van Kampen, Jorn and Salazar, Mauro and Hofman, Theo},
    pages = {1--6},
    publisher = {IEEE},
    isbn = {9798350344455},
    doi = {10.1109/VPPC60535.2023.10403195}
}

@article{Fahdzyana2022,
    title = {{Decomposition-Based Integrated Optimal Electric Powertrain Design}},
    year = {2022},
    journal = {IEEE Transactions on Vehicular Technology},
    author = {Fahdzyana, Chyannie Amarillio and Salazar, Mauro and Donkers, M. C.F. and Hofman, Theo},
    number = {6},
    pages = {6044--6058},
    volume = {71},
    publisher = {IEEE},
    doi = {10.1109/TVT.2022.3156472},
    issn = {19399359}
}

@article{Optimization,
    title = {{Outline Chapter II}},
    author = {Optimization, Winding},
    pages = {9--11}
}

@article{Silvas2016,
    title = {{Review of Optimization Strategies for System-Level Design in Hybrid Electric Vehicles}},
    year = {2016},
    journal = {IEEE Transactions on Vehicular Technology},
    author = {Silvas, Emilia and Hofman, Theo and Murgovski, Nikolce and Etman, L. F.Pascal and Steinbuch, Maarten},
    number = {1},
    pages = {57--70},
    volume = {66},
    publisher = {IEEE},
    doi = {10.1109/TVT.2016.2547897},
    issn = {19399359}
}

@article{Deb2002ANSGA-II,
    title = {{A fast and elitist multiobjective genetic algorithm: NSGA-II}},
    year = {2002},
    journal = {IEEE Transactions on Evolutionary Computation},
    author = {Deb, K. and Pratap, A. and Agarwal, S. and Meyarivan, T.},
    number = {2},
    month = {4},
    pages = {182--197},
    volume = {6},
    url = {http://ieeexplore.ieee.org/document/996017/},
    doi = {10.1109/4235.996017},
    issn = {1089778X}
}

@article{R.1980ASplines.,
    title = {{A Practical Guide to Splines.}},
    year = {1980},
    journal = {Mathematics of Computation},
    author = {R., J. and de Boor, Carl},
    number = {149},
    month = {1},
    pages = {325},
    volume = {34},
    publisher = {JSTOR},
    doi = {10.2307/2006241},
    issn = {00255718}
}

@article{Yang2022AAccuracy,
    title = {{A shape preserving C2 non-linear, non-uniform, subdivision scheme with fourth-order accuracy}},
    year = {2022},
    journal = {Applied and Computational Harmonic Analysis},
    author = {Yang, Hyoseon and Yoon, Jungho},
    number = {3},
    month = {9},
    pages = {267--292},
    volume = {60},
    publisher = {Academic Press},
    url = {https://doi.org/10.1007/s00211-016-0809-y},
    doi = {10.1016/j.acha.2022.03.006},
    issn = {1096603X}
}

@article{Lin2014AVehicle,
    title = {{A Traction Control Strategy with an Efficiency Model in a Distributed Driving Electric Vehicle}},
    year = {2014},
    journal = {The Scientific World Journal},
    author = {Lin, Cheng and Cheng, Xingqun},
    pages = {261085},
    volume = {2014},
    publisher = {Hindawi Publishing Corporation},
    url = {https://pmc.ncbi.nlm.nih.gov/articles/PMC4146358/},
    doi = {10.1155/2014/261085},
    issn = {1537744X},
    pmid = {25197697}
}

@article{Peddada2021AnOpportunities,
    title = {{An Introduction to 3D SPI2 (Spatial Packaging of Interconnected Systems With Physics Interactions) Design Problems: A Review of Related Work, Existing Gaps, Challenges, and Opportunities}},
    year = {2021},
    journal = {Proceedings of the ASME Design Engineering Technical Conference},
    author = {Peddada, Satya R.T. and Zeidner, Lawrence E. and James, Kai A. and Allison, James T.},
    month = {11},
    volume = {3B-2021},
    publisher = {American Society of Mechanical Engineers Digital Collection},
    url = {https://dx.doi.org/10.1115/DETC2021-72106},
    isbn = {9780791885390},
    doi = {10.1115/DETC2021-72106}
}

@article{Kim2003AnalyticalDesign,
    title = {{Analytical Target Cascading in Automotive Vehicle Design}},
    year = {2003},
    journal = {Journal of Mechanical Design},
    author = {Kim, Hyung Min and Rideout, D. Geoff and Papalambros, Panos Y. and Stein, Jeffrey L.},
    number = {3},
    month = {9},
    pages = {481--489},
    volume = {125},
    publisher = {American Society of Mechanical Engineers Digital Collection},
    url = {https://dx.doi.org/10.1115/1.1586308},
    doi = {10.1115/1.1586308},
    issn = {1050-0472}
}

@inproceedings{Shoemake1985AnimatingCurves,
    title = {{Animating Rotation with Quaternion Curves}},
    year = {1985},
    booktitle = {SIGGRAPH '85},
    author = {Shoemake, Ken},
    number = {3},
    month = {7},
    pages = {245--254},
    volume = {19},
    url = {https://dl.acm.org/doi/abs/10.1145/325334.325242},
    address = {San Fransisco},
    doi = {10.1145/325334.325242}
}

@article{Sapietova2018ApplicationDesigning,
    title = {{Application of optimization algorithms for robot systems designing}},
    year = {2018},
    journal = {International Journal of Advanced Robotic Systems},
    author = {Sapietov{\'{a}}, Alžbeta and S{\'{a}}ga, Milan and Kuric, Ivan and V{\'{a}}clav, Štefan},
    number = {1},
    month = {1},
    volume = {15},
    publisher = {SAGE Publications Inc.},
    url = {https://journals.sagepub.com/doi/full/10.1177/1729881417754152},
    doi = {10.1177/1729881417754152;WEBSITE:WEBSITE:SAGE;WGROUP:STRING:PUBLICATION},
    issn = {17298814}
}

@article{vanKampen2026AutomatedConstraints,
    title = {{Automated Three-Dimensional Spatial Optimization for Multidomain Systems With Alignment Constraints}},
    year = {2026},
    journal = {Journal of Mechanical Design},
    author = {van Kampen, Jorn and Salazar, Mauro and Hofman, Theo},
    number = {5},
    month = {5},
    volume = {148},
    publisher = {ASME International},
    url = {https://dx.doi.org/10.1115/1.4070618},
    doi = {10.1115/1.4070618},
    issn = {1050-0472}
}

@article{Belingardi2023BatterySolutions,
    title = {{Battery Pack and Underbody: Integration in the Structure Design for Battery Electric Vehicles—Challenges and Solutions}},
    year = {2023},
    journal = {Vehicles 2023, Vol. 5, Pages 498-514},
    author = {Belingardi, Giovanni and Scattina, Alessandro},
    number = {2},
    month = {4},
    pages = {498--514},
    volume = {5},
    publisher = {Multidisciplinary Digital Publishing Institute},
    url = {https://www.mdpi.com/2624-8921/5/2/28},
    doi = {10.3390/vehicles5020028},
    issn = {26248921}
}

@article{Lagarias1998ConvergenceDimensions,
    title = {{Convergence Properties of the Nelder--Mead Simplex Method in Low Dimensions}},
    year = {1998},
    journal = {SIAM Journal on Optimization},
    author = {Lagarias, Jeffrey C and Reeds, James A and Wright, Margaret H and Wright, Paul E},
    number = {1},
    month = {1},
    pages = {112--147},
    volume = {9},
    publisher = {Society for Industrial and Applied Mathematics},
    url = {https://doi.org/10.1137/S1052623496303470},
    doi = {10.1137/S1052623496303470},
    issn = {1052-6234}
}

@article{Bayrak2016Decomposition-BasedDesign,
    title = {{Decomposition-Based Design Optimization of Hybrid Electric Powertrain Architectures: Simultaneous Configuration and Sizing Design}},
    year = {2016},
    journal = {Journal of Mechanical Design},
    author = {Bayrak, Alparslan Emrah and Kang, Namwoo and Papalambros, Panos Y.},
    number = {7},
    month = {7},
    volume = {138},
    publisher = {American Society of Mechanical Engineers (ASME)},
    url = {https://dx.doi.org/10.1115/1.4033655},
    doi = {10.1115/1.4033655/376177},
    issn = {10500472}
}

@article{Tutuianu2015DevelopmentLegislation,
    title = {{Development of the World-wide harmonized Light duty Test Cycle (WLTC) and a possible pathway for its introduction in the European legislation}},
    year = {2015},
    journal = {Transportation Research Part D: Transport and Environment},
    author = {Tutuianu, Monica and Bonnel, Pierre and Ciuffo, Biagio and Haniu, Takahiro and Ichikawa, Noriyuki and Marotta, Alessandro and Pavlovic, Jelica and Steven, Heinz},
    month = {10},
    pages = {61--75},
    volume = {40},
    publisher = {Pergamon},
    url = {https://www.sciencedirect.com/science/article/pii/S1361920915001030},
    doi = {10.1016/J.TRD.2015.07.011},
    issn = {1361-9209}
}

@techreport{Rosso2021EfficientVehicles,
    title = {{Efficient Design of Integrated Underbody and Battery Pack for Battery Electric Vehicles}},
    year = {2021},
    author = {Rosso, Gabriele},
    month = {10},
    pages = {199},
    url = {https://webthesis.biblio.polito.it/20128/1/tesi.pdf},
    institution = {Politecnico Di Torino},
    address = {}
}

@article{Verbruggen2020ElectricTrucks,
    title = {{Electric Powertrain Topology Analysis and Design for Heavy-Duty Trucks}},
    year = {2020},
    journal = {Energies},
    author = {Verbruggen, Frans and Hofman, Theo and Silvas, Emilia},
    number = {10},
    pages = {2434},
    volume = {13},
    doi = {10.3390/en13102434}
}

@incollection{Bauchau2009Euler-BernoulliTheory,
    title = {{Euler-Bernoulli beam theory}},
    year = {2009},
    booktitle = {Structural Analysis},
    author = {Bauchau, O A and Craig, J I},
    editor = {Bauchau, O A and Craig, J I},
    pages = {173--221},
    publisher = {Springer Netherlands},
    url = {https://doi.org/10.1007/978-90-481-2516-6_5},
    address = {Dordrecht},
    isbn = {978-90-481-2516-6},
    doi = {10.1007/978-90-481-2516-6{\_}5}
}

@article{Ehrgott2026FiftyComputation,
    title = {{Fifty years of multi-objective optimization and decision-making: From mathematical programming to evolutionary computation}},
    year = {2026},
    journal = {European Journal of Operational Research},
    author = {Ehrgott, Matthias and K{\"{o}}ksalan, Murat and Kadzi{\'{n}}ski, Miłosz and Deb, Kalyanmoy},
    number = {1},
    month = {4},
    pages = {1--25},
    volume = {330},
    publisher = {North-Holland},
    url = {https://www.sciencedirect.com/science/article/pii/S0377221725004849},
    doi = {10.1016/J.EJOR.2025.06.012},
    issn = {0377-2217}
}

@techreport{Westerhof2025HybridSystems.,
    title = {{Hybrid Optimization for Spatial Packaging of Interconnected Systems.}},
    year = {2025},
    author = {Westerhof, Steven and Hofman, Theo and Van Kampen, Jorn},
    month = {12},
    url = {https://research.tue.nl/en/studentTheses/hybrid-optimization-for-spatial-packaging-of-interconnected-syste/},
    institution = {TU Eindhoven},
    address = {Eindhoven},
    keywords = {Maximal Disjoint Ball Decomposition (MDBD), Spatial Packaging of Interconnected Systems{\\\}with Physical Interactions (SPI2), generative design, hybrid{\\\}optimization, placement and routing optimization}
}

@article{Danielsson2016InfluenceCharacteristics,
    title = {{Influence of body stiffness on vehicle dynamics characteristics}},
    year = {2016},
    journal = {The Dynamics of Vehicles on Roads and Tracks - Proceedings of the 24th Symposium of the International Association for Vehicle System Dynamics, IAVSD 2015},
    author = {Danielsson, O. and Gonz{\'{a}}lez Coca{\~{n}}a, A. and Ekstr{\"{o}}m, K. and Bayani Khaknejad, M. and Klomp, M. and Dekker, R.},
    pages = {45--56},
    publisher = {CRC Press/Balkema},
    isbn = {9781138028852},
    doi = {10.1201/B21185-7}
}

@article{Schuman2005IntegratedModeling,
    title = {{Integrated system-level optimization for concurrent engineering with parametric subsystem modeling}},
    year = {2005},
    journal = {Collection of Technical Papers - AIAA/ASME/ASCE/AHS/ASC Structures, Structural Dynamics and Materials Conference},
    author = {Schuman, Todd and De Weck, Olivier L. and Sobieski, Jaroslaw},
    pages = {4989--5008},
    volume = {7},
    url = {/doi/pdf/10.2514/6.2005-2199},
    doi = {10.2514/6.2005-2199},
    issn = {02734508}
}

@article{Chen2020MaximalAnalysis,
    title = {{Maximal Disjoint Ball Decompositions for shape modeling and analysis}},
    year = {2020},
    journal = {Computer-Aided Design},
    author = {Chen, Jiangce and Ilie{\c{s}}, Horea T.},
    number = {4},
    month = {9},
    pages = {102850},
    volume = {126},
    publisher = {Elsevier},
    url = {https://doi.org/10.1080/16864360.2016.1257192},
    doi = {10.1016/j.cad.2020.102850},
    issn = {00104485}
}

@book{Fenner2012MechanicsStructures,
    title = {{Mechanics of solids and structures}},
    year = {2012},
    author = {Fenner, Roger T.. and Reddy, J. N..},
    pages = {683},
    publisher = {CRC Press/Taylor {\&} Francis Group},
    isbn = {9781439858141}
}

@article{Pavllo2019ModelingNetworks,
    title = {{Modeling Human Motion with Quaternion-based Neural Networks}},
    year = {2019},
    journal = {International Journal of Computer Vision},
    author = {Pavllo, Dario and Feichtenhofer, Christoph and Auli, Michael and Grangier, David},
    number = {4},
    month = {10},
    pages = {855--872},
    volume = {128},
    publisher = {Springer},
    url = {http://arxiv.org/abs/1901.07677 http://dx.doi.org/10.1007/s11263-019-01245-6},
    doi = {10.1007/s11263-019-01245-6},
    arxivId = {1901.07677}
}

@article{Othaganont2017Multi-objectiveTopologies,
    title = {{Multi-objective optimisation for battery electric vehicle powertrain topologies}},
    year = {2017},
    journal = {Proceedings of the Institution of Mechanical Engineers, Part D: Journal of Automobile Engineering},
    author = {Othaganont, Pongpun and Assadian, Francis and Auger, Daniel J.},
    number = {8},
    month = {7},
    pages = {1046--1065},
    volume = {231},
    publisher = {SAGE Publications Ltd},
    url = {/doi/pdf/10.1177/0954407016671275?download=true},
    doi = {10.1177/0954407016671275;ISSUE:ISSUE:DOI},
    issn = {20412991}
}

@article{Hofstetter2018Multi-ObjectiveDevelopment,
    title = {{Multi-Objective System Design Synthesis for Electric Powertrain Development}},
    year = {2018},
    journal = {2018 IEEE Transportation and Electrification Conference and Expo, ITEC 2018},
    author = {Hofstetter, Martin and Hirz, Mario and Gintzel, Martin and Schmidhofer, Andreas},
    month = {8},
    pages = {274--279},
    publisher = {Institute of Electrical and Electronics Engineers Inc.},
    url = {https://ieeexplore.ieee.org/document/8450113},
    isbn = {9781538630488},
    doi = {10.1109/ITEC.2018.8450113}
}

@article{Cohen1978OptimizationApproach,
    title = {{Optimization by Decomposition and Coordination: A Unified Approach}},
    year = {1978},
    journal = {IEEE Transactions on Automatic Control},
    author = {Cohen, Guy},
    number = {2},
    pages = {222--232},
    volume = {23},
    url = {https://ieeexplore.ieee.org/document/1101718},
    doi = {10.1109/TAC.1978.1101718},
    issn = {15582523}
}

@article{Audet2021PerformanceOptimization,
    title = {{Performance indicators in multiobjective optimization}},
    year = {2021},
    journal = {European Journal of Operational Research},
    author = {Audet, Charles and Bigeon, Jean and Cartier, Dominique and Le Digabel, Sébastien and Salomon, Ludovic},
    number = {2},
    month = {7},
    pages = {397--422},
    volume = {292},
    publisher = {North-Holland},
    url = {https://www.sciencedirect.com/science/article/pii/S0377221720309620},
    doi = {10.1016/J.EJOR.2020.11.016},
    issn = {0377-2217}
}

@article{Sola2017QuaternionFilter,
    title = {{Quaternion kinematics for the error-state Kalman filter}},
    year = {2017},
    journal = {CoRR},
    author = {Sol{\`{a}}, Joan},
    month = {11},
    volume = {abs/1711.02508},
    url = {http://arxiv.org/abs/1711.02508},
    arxivId = {1711.02508}
}

@article{Diebel2006RepresentingVectors,
    title = {{Representing Attitude: Euler Angles, Unit Quaternions, and Rotation Vectors}},
    year = {2006},
    journal = {Matrix},
    author = {Diebel, James},
    pages = {1--35},
    volume = {58},
    url = {https://www.researchgate.net/publication/215458871_Representing_Attitude_Euler_Angles_Unit_Quaternions_and_Rotation_Vectors}
}

@article{Young1976RoarksStrain,
    title = {{Roark's Formulas for Stress and Strain}},
    year = {1976},
    author = {Young, Warren C and Budynas, Richard G and York, New and San, Chicago and Lisbon, Francisco and Madrid, London and City, Mexico and New, Milan and San, Delhi and Seoul, Juan},
    isbn = {007072542X}
}

@article{Behzadi2025SpatialSystems,
    title = {{Spatial Packaging and Routing Optimization of Complex Interacting Engineered Systems}},
    year = {2025},
    journal = {Journal of Mechanical Design},
    author = {Behzadi, Mohammad M. and Zaffetti, Peter and Chen, Jiangce and Zeidner, Lawrence E. and Ilies, Horea T.},
    number = {7},
    month = {7},
    volume = {147},
    publisher = {American Society of Mechanical Engineers Digital Collection},
    url = {https://dx.doi.org/10.1115/1.4067427},
    doi = {10.1115/1.4067427},
    issn = {1050-0472}
}

@article{Levickas2025StabilityBraking,
    title = {{Stability Issues of Rear–Wheel–Drive Electric Vehicle During Regenerative Braking}},
    year = {2025},
    journal = {Applied Sciences 2025, Vol. 15,},
    author = {Levickas, Rapolas and {\v{Z}}uraulis, Vidas},
    number = {20},
    month = {10},
    volume = {15},
    publisher = {Multidisciplinary Digital Publishing Institute},
    url = {https://www.mdpi.com/2076-3417/15/20/10926},
    doi = {10.3390/app152010926},
    issn = {20763417}
}

@article{Lu2023TopologyBatteries,
    title = {{Topology optimization of electric vehicle chassis structure with distributed load-bearing batteries}},
    year = {2023},
    journal = {Structural and Multidisciplinary Optimization 2023 66:6},
    author = {Lu, Yufan and Mao, Hongjiang and Zhou, Mingdong},
    number = {6},
    month = {5},
    pages = {134-},
    volume = {66},
    publisher = {Springer},
    url = {https://link.springer.com/article/10.1007/s00158-023-03578-w},
    isbn = {0123456789},
    doi = {10.1007/s00158-023-03578-w},
    issn = {16151488}
}

@article{Peddada2022TowardSPI2,
    title = {{Toward Holistic Design of Spatial Packaging of Interconnected Systems with Physical Interactions (SPI2)}},
    year = {2022},
    journal = {Journal of Mechanical Design},
    author = {Peddada, Satya R.T. and Zeidner, Lawrence E. and Ilies, Horea T. and James, Kai A. and Allison, James T.},
    number = {12},
    month = {12},
    volume = {144},
    publisher = {American Society of Mechanical Engineers (ASME)},
    url = {https://dx.doi.org/10.1115/1.4055055},
    doi = {10.1115/1.4055055/1143331},
    issn = {10500472}
}

@book{Guzzella2007VehicleOptimization,
    title = {{Vehicle propulsion systems: Introduction to modeling and optimization}},
    year = {2007},
    booktitle = {Vehicle Propulsion Systems (Second Edition): Introduction to Modeling and Optimization},
    author = {Guzzella, Lino and Sciarretta, Antonio},
    pages = {1--338},
    publisher = {Springer Berlin Heidelberg},
    isbn = {9783540746911},
    doi = {10.1007/978-3-540-74692-8/COVER}
}
